\documentclass[authoryear,preprint,12pt]{article}

\usepackage{amssymb}
\usepackage[utf8]{inputenc} 
\usepackage[english]{babel} 
\usepackage{microtype} 
\usepackage{natbib} 
\usepackage[hidelinks]{hyperref} 
\usepackage[margin=5pt,skip=0pt]{caption} 
\usepackage[margin=5pt,skip=0pt]{subcaption}
\usepackage[shortlabels]{enumitem}
\usepackage{placeins}
\usepackage{wrapfig}
\usepackage{url}
\usepackage{amsmath}
\usepackage{color}
\usepackage[percent]{overpic}
\usepackage{graphicx}
\usepackage{authblk}

\graphicspath{{imgs/}}

\definecolor{mygray}{gray}{0.4}

\begin{document}

\title{Unsteady aerodynamic effects in small-amplitude pitch oscillations of an airfoil}

\author{P. S. Negi\footnote{negi@mech.kth.se}}
\author{R. Vinuesa}
\author{A. Hanifi}
\author{P. Schlatter}
\author{D. S. Henningson}
\affil{\small Department of Mechanics, Linn\'e FLOW Centre and Swedish e-Science Research Centre (SeRC), KTH Royal Institute of Technology, SE-100 44 Stockholm, Sweden}
\date{}

\maketitle

\begin{abstract}
	High-fidelity wall-resolved large-eddy simulations (LES) are utilized to investigate the flow-physics of small-amplitude pitch oscillations of an airfoil at $Re_{c}=100,000$. The investigation of the unsteady phenomenon is done in the context of natural laminar flow airfoils, which can display sensitive dependence of the aerodynamic forces on the angle of attack in certain ``off-design'' conditions. The dynamic range of the pitch oscillations is chosen to be in this sensitive region. Large variations of the transition point on the suction-side of the airfoil are observed throughout the pitch cycle resulting in a dynamically rich flow response. Changes in the stability characteristics of a leading-edge laminar separation bubble has a dominating influence on the boundary layer dynamics and causes an abrupt change in the transition location over the airfoil. The LES procedure is based on a relaxation-term which models the dissipation of the smallest unresolved scales. The validation of the procedure is provided for channel flows and for a stationary wing at $Re_{c}=400,000$.
\end{abstract}
\textbf{Keywords - } unsteady aerodynamics, dynamic-response, transition, wall-resolved les, laminar separation bubble, local stability

\section{Introduction}

A large focus of the research on unsteady wings tends towards stall dynamics such as the earlier works of \cite{mccroskey81,mccroskey82experimental,mccroskey73,mccroskey76,carr1977,ericsson_stall88a,ericsson_stall88b}, \emph{etc.} More recent works by \cite{dunne2015,rival2010,choudhry14,visbal11,visbal14,visbal17,alferez13,rosti16} \emph{etc.} continue the investigation which appears far from complete. The review by \cite{mccroskey82} and a more recent one by \cite{coorke15} provide an overview of the dynamic stall process. Studies on the aerodynamic behavior of small-amplitude pitch oscillations are typically done from the perspective of aeroelasticity where investigations tend to focus on the time varying aerodynamic forces on the airfoil with much less attention paid to the boundary-layer characteristics. Some studies focusing on the time dependent boundary layer in small pitch amplitudes include the work done by \cite{pascazio96} which shows a time delay in laminar-turbulent transition during pitching. \cite{nati15} analyzed the effect of small amplitude pitching on a laminar separation bubble at low Reynolds numbers. \cite{mai11} and \cite{hebler13} investigate the boundary-layer changes in an oscillating natural laminar flow airfoil in transonic conditions.
\begin{figure}[h]
	\centering
	\includegraphics[width=0.95\textwidth]{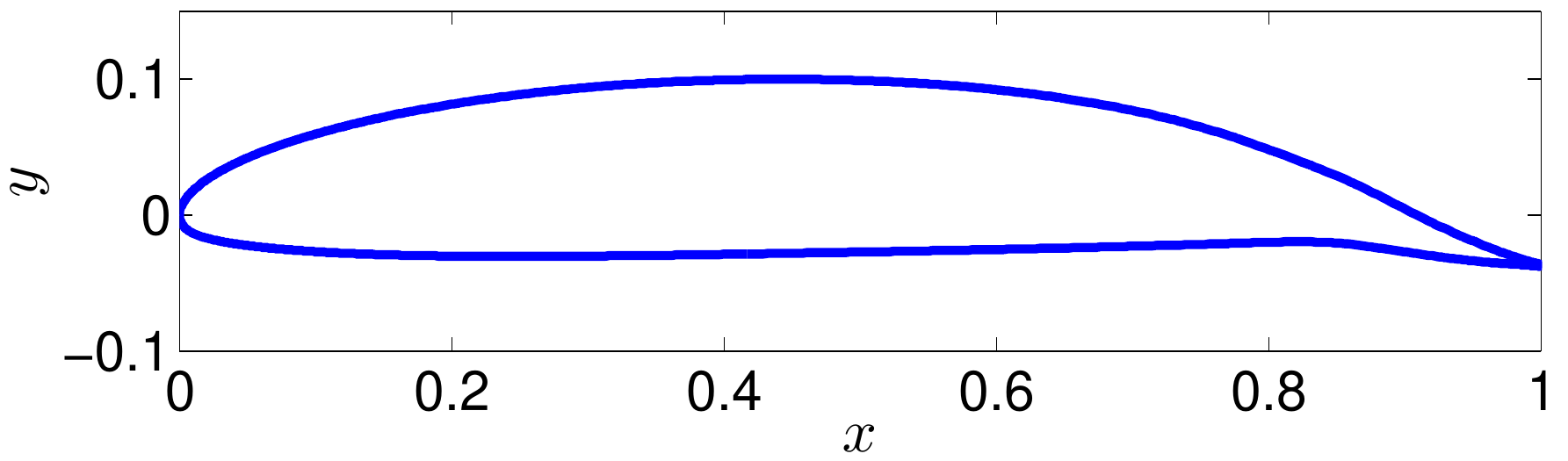}
	\caption{Natural Laminar Flow (NLF) airfoil tested at the Aeronautical and Vehicle engineering department of KTH \citep{lokatt17,lokattthesis}}
	\label{fig:foil_david}
\end{figure}
Such cases qualitatively represent small changes in operating conditions, such as the changes due to structural deformations or small trailing-edge flap deflections. The understanding of flow response to such changes can be crucial in cases where small perturbations induce large variations in aerodynamic forces. Such sensitive dependence of aerodynamic forces may be found in the static characteristics of natural laminar flow (NLF) airfoils for a certain range of angle of attack. Their performance critically depends on maintaining laminar flow over the suction side of the airfoil and a loss of laminar flow over the airfoil causes large variations of the aerodynamic forces. In addition to such sensitive dependence of the aerodynamic characteristics, recent transonic unsteady experiments using NLF airfoils have brought to light a peculiar property of these airfoils. The unsteady aerodynamic coefficients for laminar airfoils exhibit a non-linear dynamic response to simple harmonic pitch motions \citep{mai11,hebler13}. Such a non-linear response is inconsistent with the predictions of classical unsteady aerodynamic models \citep{theodorsen35} which are typically based on inviscid assumptions. Similar experiments within the subsonic range have been performed by \cite{lokattthesis} who also found strongly non-linear behavior of the normal force coefficient. These non-linearities occur only for oscillations within a certain range of angle of attack ($\alpha$) and have been strongly linked to the free movement of transition over the suction side of the airfoil. They seem to be nearly absent when suction side transition is fixed at the leading-edge \citep{mai11,lokattthesis}. =While the above experiments were performed at a Reynolds number range of $O(10^{6})$, simulations and experiments at lower transitional Reynolds numbers on a NACA 0012 airfoil have also been shown to exhibit non-linear unsteady aerodynamic characteristics \citep{poirel08,poirel10,barnes16}. At these transitional Reynolds numbers, the laminar separation bubble has been shown to play a key role in the unsteady dynamics \citep{yuan13,barnes18}. In all the above cases, the changes in boundary-layer characteristics lead to dynamic phenomena which can not be modeled using linear theories. With these effects in mind, the current work seeks to investigate the flow dynamics of an unsteady airfoil within the aerodynamic regime where large changes in the boundary-layer characteristics are observed. The analysis of the unsteady laminar separation bubble which develops near the leading-edge is done using local linear stability analysis in order to explain some of the observed unsteady dynamics.

\begin{figure}[h]
	\centering
	\includegraphics[width=0.65\textwidth]{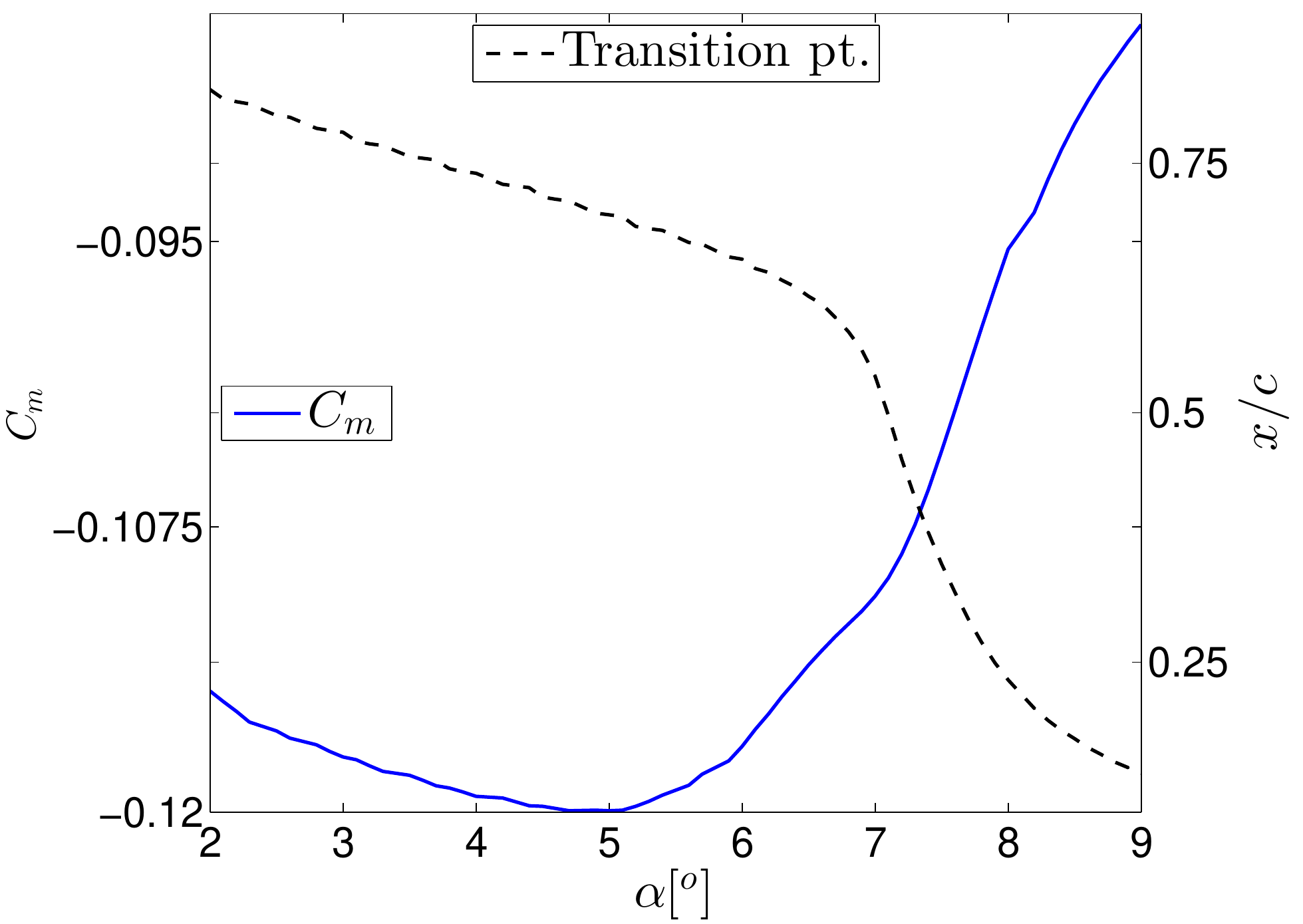}
	\caption{Coefficient of moment ($C_{m}$) (values on the left axis), and suction side transition location (values on the right axis) as calculated using XFOIL.}
	\label{fig:xfoil_cm}
\end{figure}

The present work investigates the flow physics of small-amplitude pitch oscillations of a laminar airfoil (figure~\ref{fig:foil_david}). The airfoil was designed at the Aeronautical and Vehicle Engineering department of KTH, where it has been used in previous experimental and numerical works \citep{lokatt17}. The same airfoil was used in the unsteady experiments of \cite{lokattthesis}. The simulations are performed at a chord-based Reynolds number of $Re_{c}=100,000$. The angle of attack range for the oscillation was chosen from the static characteristics of the airfoil. The static characteristics were calculated using an integral boundary layer code XFOIL \citep{drela89}, which predicted sharp changes in the coefficient of moment ($C_{m}$) and suction-side transition location (figure~\ref{fig:xfoil_cm}) above an angle of attack $\alpha>6^{\circ}$. The steep slope of the coefficient of moment curve indicates a region where aerodynamic forces are sensitive to small changes in $\alpha$. It is important to note that the results obtained from XFOIL are only used to approximately identify the above mentioned region of sensitive dependence for the airfoil at a low computational cost. Simulations with a static airfoil are performed to ensure that such sensitive dependence on angle of attack is indeed observed with the high-fidelity LES. The exact dynamic range of the pitching motion is prescribed based on the results of the static simulations (described in section 3).

In recent works, wall-resolved large-eddy simulations have proven to be an effective tool for studying flow physics at high Reynolds numbers with a computational cost which is much lower than that of direct numerical simulations (DNS). Some of the works to utilize this method include spatially evolving boundary layers \citep{eitel14}, turbulent channel flows \citep{schlatter06b,schlatter06}, pipe flows \citep{chin15} and flow over wings \citep{uzun10,lombard15}. Successful application of the approach has motivated its use in the present work, which aims to gain insight into the flow-physics of unsteady airfoils undergoing small amplitude pitch oscillations.

The remainder of the paper is divided into 3 sections. Section 2 describes the numerical setup and presents the results of the validation of the LES. Results of both the stationary and pitching simulations are discussed in Section 3. The conclusions of the study are presented in Section 4.

\section{Computational setup}

\subsection{Numerical method}

The computational code used for the simulations is Nek5000, which is an open-source incompressible Navier--Stokes solver developed by \cite{nek5000} at Argonne National Laboratory. It is a based on a spectral-element method which allows the mapping of elements to complex geometries along with a high-order spatial discretization within the elements. The method uses Lagrange interpolants as basis functions and utilizes Gauss--Lobatto--Legendre (GLL) quadrature for the distribution of points within the elements. The spatial discretization is done by means of the Galerkin approximation, following the $P_{N}$-$P_{N-2}$ formulation \citep{maday89}. An $11^{th}$ order polynomial approximation is used for the velocity with a $9^{th}$ order approximation for pressure. The nonlinear terms are treated explicitly by third-order extrapolation (EXT3), whereas the viscous terms are treated implicitly by a third-order backward differentiation scheme (BDF3). Aliasing errors are removed with the use of over-integration. The Arbitrary-Lagrangian-Eulerian (ALE) formulation is used to account for the mesh deformation due to the motion of the pitching airfoil \citep{ho90,ho91}. All equations are solved in non-dimensional units with the velocities normalized by the reference free-stream velocity $U_{0}$ and the length scales in all directions are normalized by the chord length $c$. The resultant non-dimensional time unit is given by $c/U_{0}$.
\subsection{Relaxation-term large-eddy simulation (RT-LES)}

The LES method is based on the RT3D variant of the ADM-RT approach first used by \cite{schlatter04}. The method supplements the governing equations with a dissipative term ($\chi\mathcal{H}(u)$). The equations for the resolved velocity and pressure thus read as
\begin{subequations}
	\begin{eqnarray}
		\frac{\partial u}{\partial t} + u\cdot\nabla u =  - \frac{1}{\rho}\nabla p + \frac{1}{Re}\nabla^{2}u -\chi\mathcal{H}(u), \\
		\nabla\cdot u = 0,
	\end{eqnarray}
\end{subequations}
where $\mathcal{H}$ is a high-pass spectral filter and $\chi$ is a model parameter. Together the two parameters determine the strength of the dissipative term. The method has been used in earlier studies of spatially developing boundary layers \citep{eitel14} and channel flows \citep{schlatter06}. The method has been shown to be reliable in predicting transition location and also preserving the characteristic structures which are seen in the DNS of transitional flows \citep{schlatter06}.

\begin{figure}[h]
	\begin{subfigure}[t]{0.5\textwidth}
		\centering
		\caption{Mean velocity profile.}		
		\includegraphics[width=0.95\textwidth]{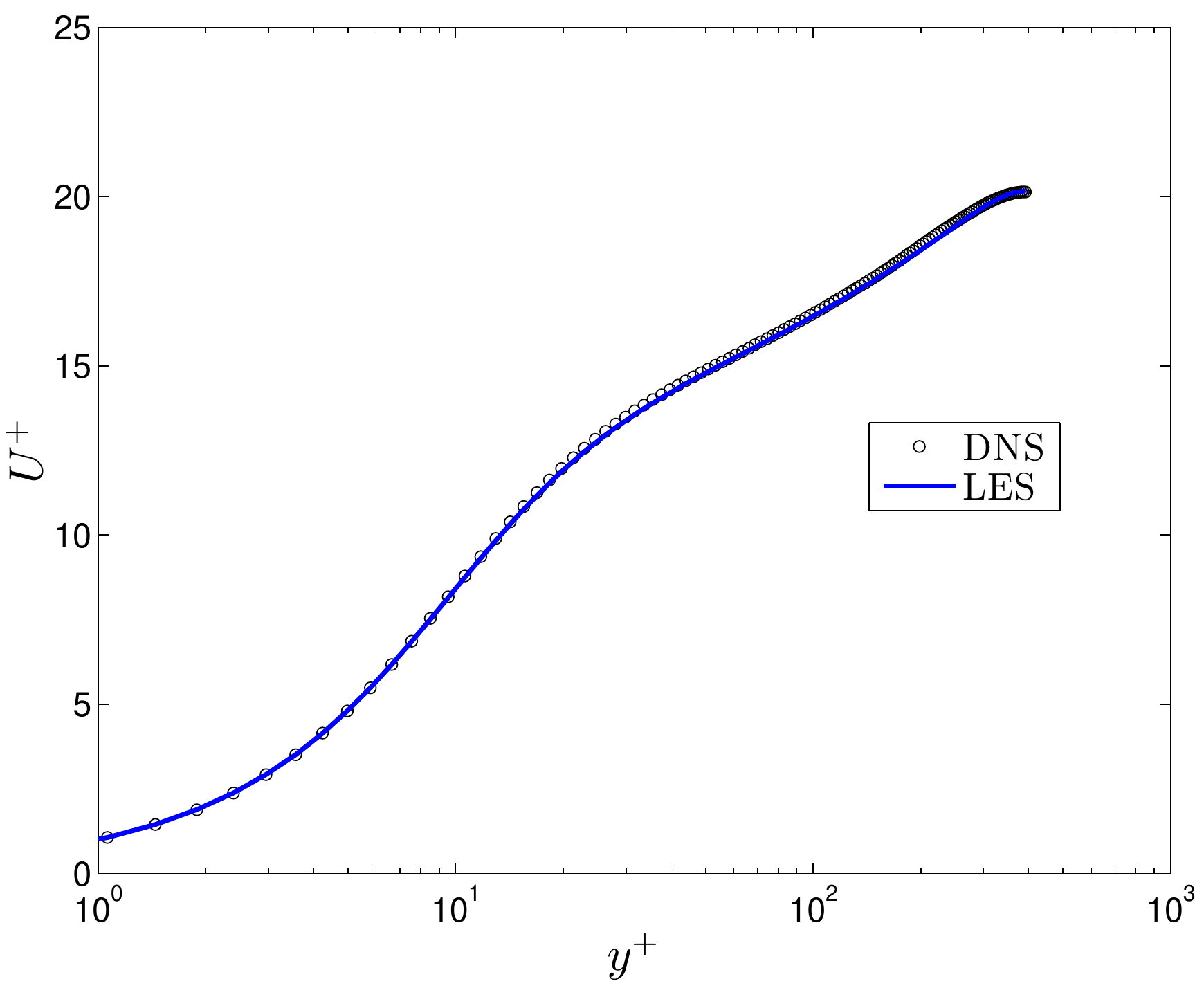}
		\label{fig:vel_mean}
	\end{subfigure}	
	\begin{subfigure}[t]{0.5\textwidth}
		\centering
		\caption{Turbulent kinetic energy budget.}		
		\includegraphics[width=0.98\textwidth]{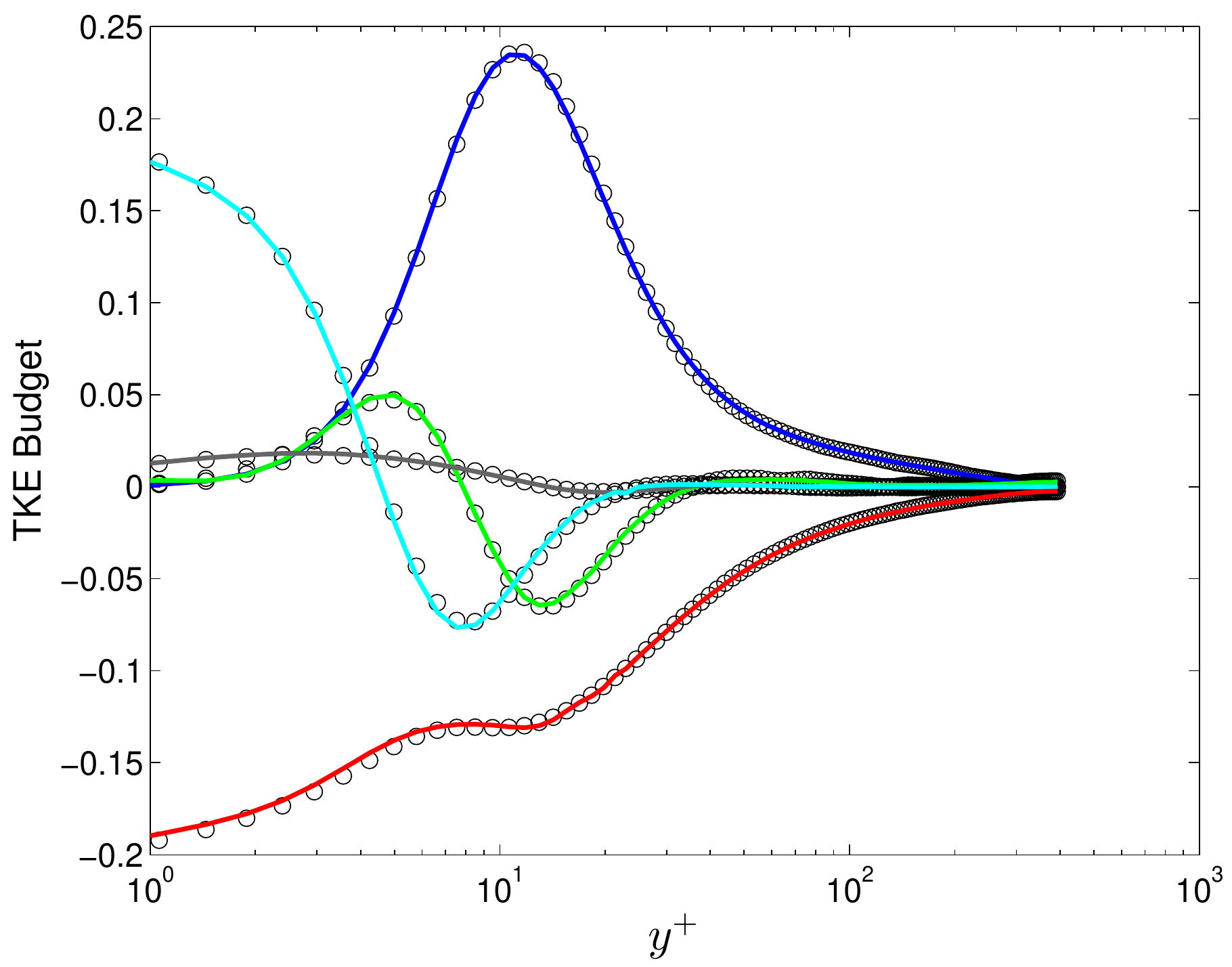}
		\label{fig:budget}
	\end{subfigure}
	\caption{Comparison of mean velocity profile and turbulent kinetic energy budget. Circles represent the DNS data from \cite{moser99} while the lines represent the values from the LES. All values are normalized with inner units. The individual terms are color coded as: Production (\textcolor{blue}{blue}), dissipation (\textcolor{red}{red}), viscous diffusion (\textcolor{cyan}{cyan}), turbulent diffusion (\textcolor{green}{green}), velocity-pressure correlation (\textcolor{mygray}{gray})}
\end{figure}

A number of tests were carried out in a channel flow at a friction Reynolds number of $Re_{\tau}=395$, and the results are compared with the DNS data of \cite{moser99}. The final mesh was set up such that the streamwise resolution was $\Delta x^{+}=18$ and the spanwise resolution was $\Delta z^{+}=9$. The first point in the wall-normal direction was set at $\Delta y_{w}^{+}=0.64$ and the wall-normal resolution near the boundary layer edge was $y_{max}^{+}=11$. The superscript $^{+}$ indicates normalization in inner units. A comparison of the results for the turbulent channel flow is shown for the mean velocity in figure~\ref{fig:vel_mean}, and for the turbulent kinetic energy (TKE) budget in figure~\ref{fig:budget}. The dissipation profile shown in the figure is the sum of resolved dissipation and the added dissipation by the relaxation term. A good agreement with the DNS is found for the mean velocity and all the kinetic energy budget terms (including the total dissipation). The resolution used this study is much finer than the typical LES and closer to the coarse DNS resolutions used in the studies of turbulent flows. A very similar resolution is used in the simulation of spatially developing boundary-layer over a flat-plate by \cite{eitel14} where the ADM-RT model is used. Similarly, LES cases of pipe flows at $Re_{\tau}\approx1000$ by \cite{chin15} uses slightly coarser resolutions than the one used in the present study.

\begin{figure}[h]
	\centering
	\includegraphics[width=0.6\textwidth]{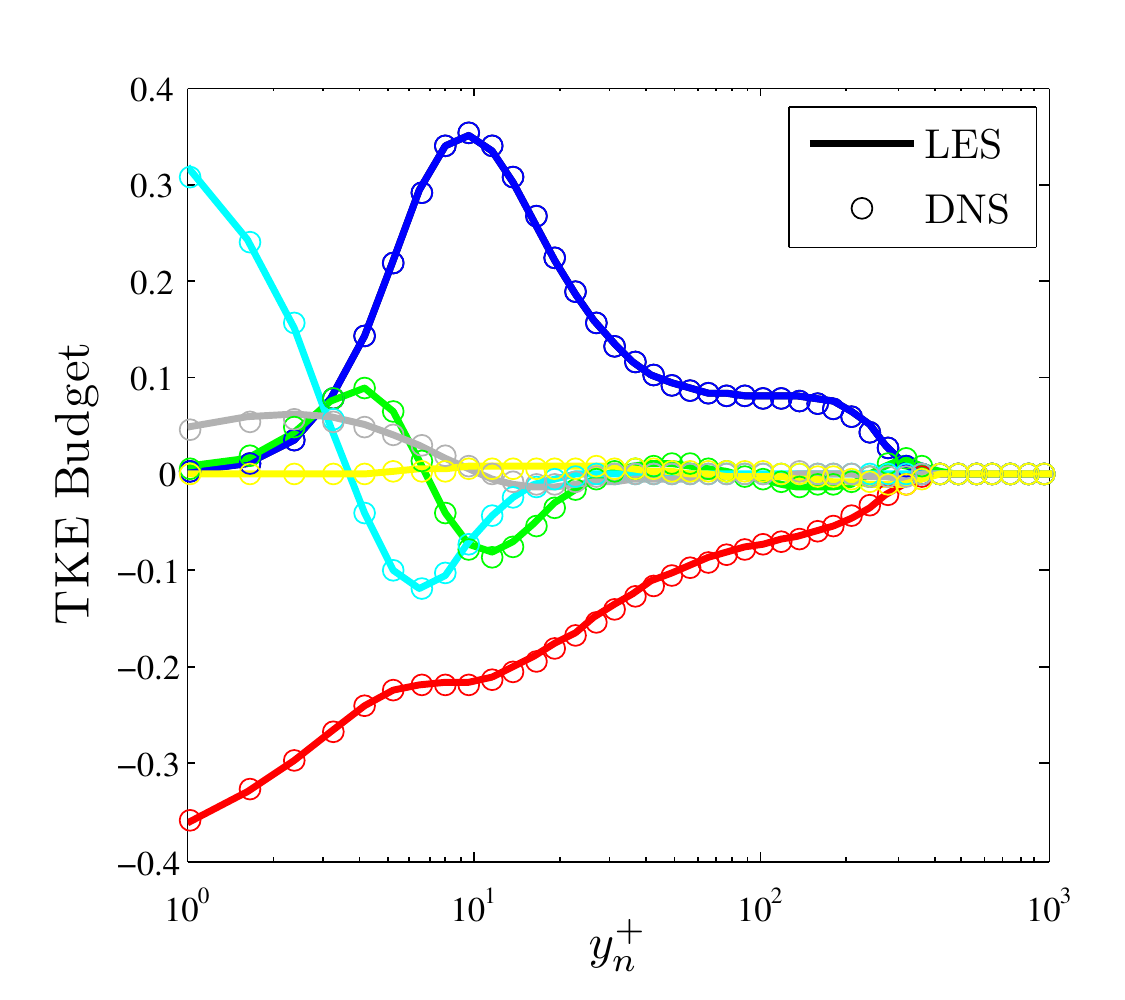}
	\caption{Comparison of turbulent kinetic energy budget for a NACA4412 wing section at the suction side location of $x/c=0.7$. The circles represent DNS data from \cite{hosseini16} while the lines are data from the LES. The individual terms are color coded as: Production (\textcolor{blue}{blue}), dissipation (\textcolor{red}{red}), viscous diffusion (\textcolor{cyan}{cyan}), turbulent diffusion (\textcolor{green}{green}), velocity-pressure correlation (\textcolor{mygray}{gray}), convection (\textcolor{yellow}{yellow})}
	\label{fig:wing_budget}
\end{figure}

\subsection{Mesh generation}

The optimum mesh resolution (in inner units) obtained in the channel flow results is then used to design the mesh around the airfoil. Wall-shear stress data is obtained using XFOIL to estimate the grid spacing on the airfoil. A trip is introduced in XFOIL at $x/c\approx0.1$ to obtain turbulent wall-shear values on both the suction and pressure sides of the airfoil. Here $c$ denotes the chord length. The grid design around the airfoil uses the following criteria:

\begin{itemize}
	\item[$\bullet$] For $0.1<x/c<0.6$, $\Delta x^{+}=18$, $\Delta y_{wall}^{+}=0.64$ and $\Delta y_{max}^{+}=11$, using the local wall-shear ($\tau_{w}$) values on the airfoil. Since the flow is expected to be laminar on the pressure side, the stream-wise resolution is slightly relaxed to $\Delta x^{+}=25$ while keeping the same wall-normal resolution.
	\item[$\bullet$] For $x/c<0.1$, the peak $\tau_{w}$ value over the suction side of the airfoil is used to estimate the grid spacing.
	\item [$\bullet$] For $x/c>0.6$, the suction side experiences a large adverse pressure gradient which significantly reduces $\tau_{w}$ values. Therefore, the $\tau_{w}$ values from the pressure side are used for both the suction and pressure sides.
	\item [$\bullet$] A structured mesh is used, which is extruded in the spanwise direction. Hence the spanwise resolution is constant throughout the domain. The resolution is set to $\Delta z^{+}=9$, where the the peak $\tau_{w}$ value from the suction side is used.
\end{itemize}

\begin{figure}[h]
	\begin{subfigure}[t]{0.49\textwidth}
		\centering
		\caption{}		
		\includegraphics[width=1.0\textwidth]{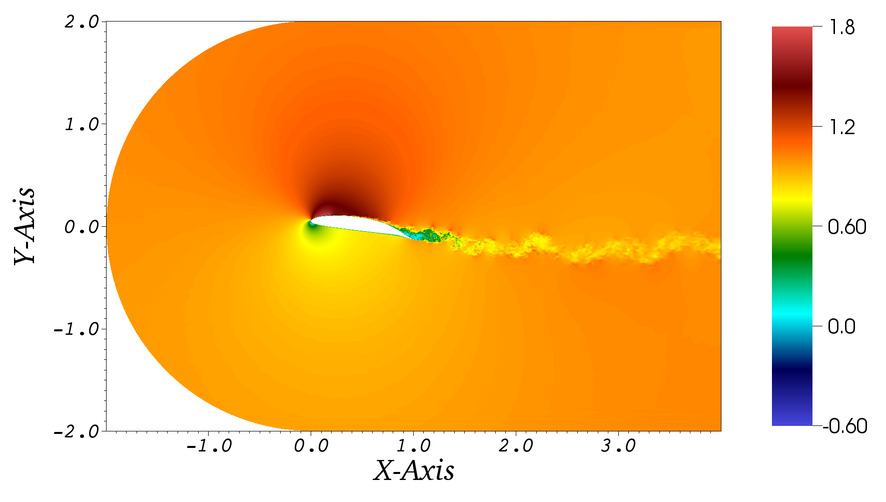}
		\label{fig:re100k_domain}
	\end{subfigure}	
	\begin{subfigure}[t]{0.49\textwidth}
		\centering
		\caption{}		
		\includegraphics[width=1.\textwidth]{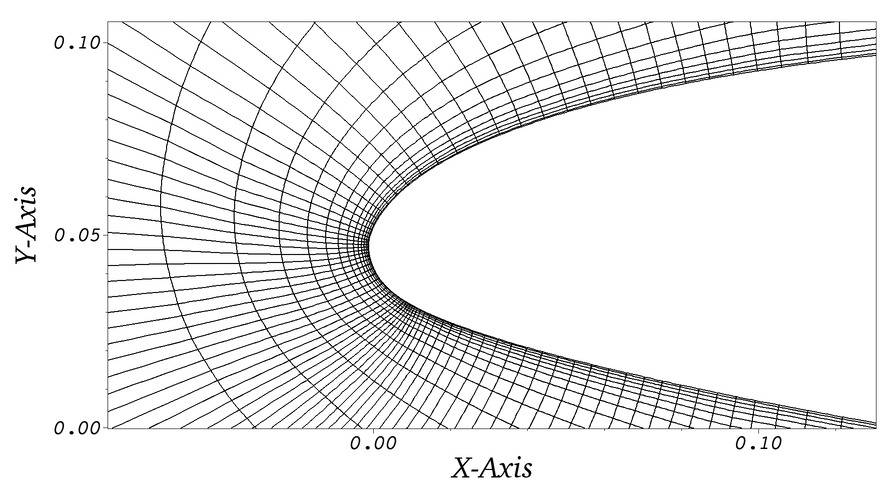}
		\label{fig:re100k_grid}
	\end{subfigure}
	\caption{(a) 2D section of the simulation domain. Colors represent the instantaneous streamwise velocity. (b) Close-up view of the spectral-element grid near the airfoil surface.}	
\end{figure}

A different criterion is needed for defining the resolution in the wake where the wall-based criteria are not valid. Accordingly, Reynolds-averaged Navier--Stokes (RANS) simulations were performed using the transition $k$-$\Omega$ SST model \citep{langtry09} with \textit{ANSYS}\textsuperscript{\textregistered} FLUENT, to estimate the Kolmogorov length scale ($\eta$) in the wake region. The RANS is setup with domain boundaries at a distance of 100 chords from the airfoil. The grid in the wake region for the LES is designed such that the average grid spacing between the GLL points follows the criteria: $\Delta x/\eta < 9$. The computational domain is set up such that the far field boundaries of the computational domain are two chords away from the airfoil leading edge in either direction. The outflow boundary is four chords downstream from the airfoil leading edge and the inlet is designed as a curved inflow boundary with a constant radial distance of two chords from the leading edge of the airfoil. The spanwise width of the domain is $l_{z}=0.25$ chords. The domain can be visualized in figure~\ref{fig:re100k_domain} and a close-up view of the spectral-elements is shown in (figure~\ref{fig:re100k_grid}). Each of the spectral-elements are further discretized by $12\times12\times12$ grid points in 3D, corresponding to an $11^{th}$ order spectral discretization. Periodic boundary conditions are imposed on the spanwise boundaries, while the energy-stabilized outflow condition suggested by \cite{dong2014} is imposed on the outflow boundary. Velocity field data for locations corresponding to the boundaries of the LES computational domain is extracted from an unsteady RANS simulation. The extracted data is imposed as a Dirichlet boundary condition on these boundaries for the LES. The method is very similar to the one used by \cite{hosseini16} in their DNS of flow around a wing section. In order to simulate low turbulence flight conditions, free-stream turbulence of intensity $Ti=0.1\%$ is superimposed on the Dirichlet boundary conditions. The free-stream turbulence is generated using Fourier modes with a von K\'arm\'an spectrum. The procedure is similar to the one described in \cite{schlatterdiploma,brandt04} and \cite{schlatter08} and has been used for the study of transition in flat plate boundary layers under the influence of free-stream turbulence. The same method has also been used for generating grid-turbulence in simulations of wind-turbines \citep{kleusberglicenciate}.

A validation of the above methodology for complex geometries such as a wing section was performed at a chord based Reynolds number of $Re_{c}=400,000$ for NACA4412 airfoil. The LES grid resolution was setup with the same grid criteria as described above. The domain boundaries and boundary conditions are identical to the setup in \cite{hosseini16}. A comparison of the wall-normal profiles of the normalized kinetic energy budget is shown in figure~\ref{fig:wing_budget}. The profiles are extracted at a streamwise location of $x/c=0.7$ on the suction side of the airfoil. The LES profiles (lines) match well with the DNS data (circles) of \cite{hosseini16}, signifying the high accuracy of the LES with the current resolution.

\section{Results and discussion}
\subsection{Steady results}
\begin{figure}[t]
	\begin{subfigure}[b]{0.49\textwidth}
		\centering
		\caption{$\alpha=6.7^{\circ}$}		
		\includegraphics[width=1\textwidth]{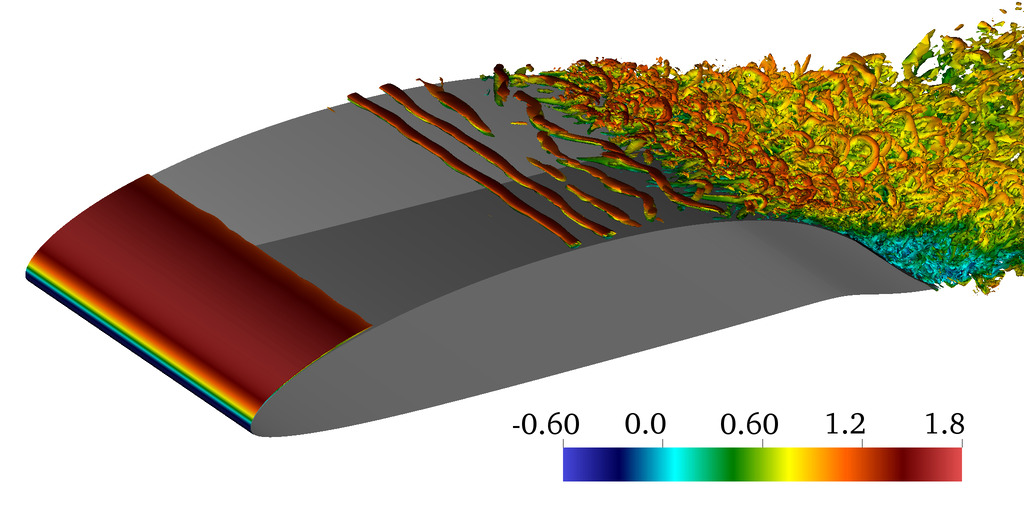}
		\label{fig:aoa67_iso}
	\end{subfigure}
	\begin{subfigure}[b]{0.49\textwidth}
		\centering
		\caption{$\alpha=8.0^{\circ}$}		
		\includegraphics[width=1\textwidth]{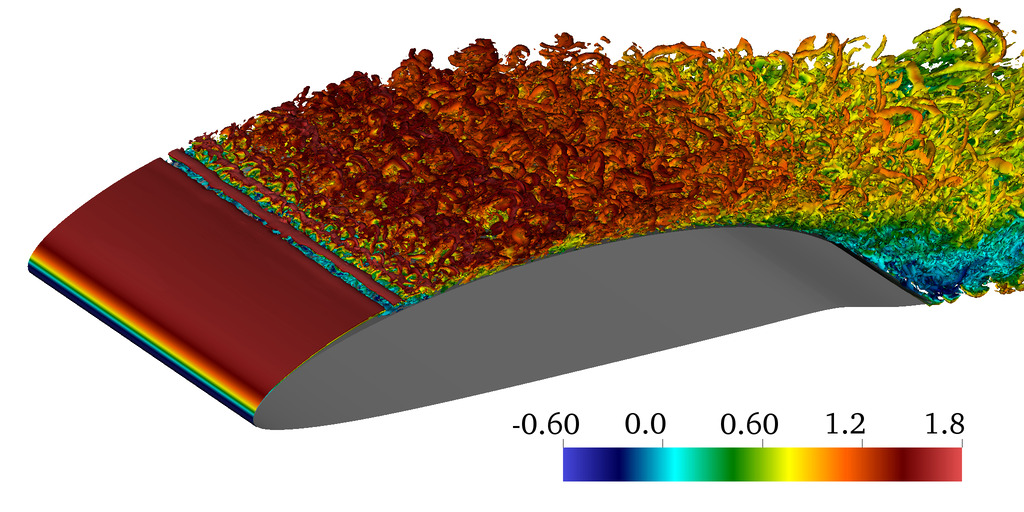}
		\label{fig:aoa80_iso}
	\end{subfigure}
	\caption{Isocontours of instantaneous $\lambda_{2}$ structures observed for two different (stationary) angles of attack.}
	\label{fig:isocontour_aoa}
\end{figure}

Simulations with a stationary airfoil were performed to investigate the location of transition without pitching motion. The simulations were performed for $Re_{c}=100,000$ at two different angles of attack ($\alpha=6.7^{\circ}$ and $\alpha=8.0^{\circ}$). As observed in figure~\ref{fig:isocontour_aoa}, the iso-contours of coherent structures, identified by negative $\lambda_{2}$ method \citep{jeong95}, show a substantial change in transition location for a small $\Delta\alpha=1.3^{\circ}$. For $\alpha=6.7^{\circ}$ the strong pressure gradient effects near the trailing edge cause transition at $x/c\approx0.7$. While for $\alpha=8.0^{\circ}$, a leading-edge laminar separation bubble forms, causing flow transition much closer to the leading edge at $x/c\approx0.2$. Such a leading-edge laminar separation bubble is not observed for the $\alpha=6.7^{\circ}$ case. The results are consistent with the trends obtained from XFOIL calculations, showing a large variation in the transition point within a small $\alpha$ change (figure~\ref{fig:xfoil_cm}).
\subsection{Unsteady boundary layer characteristics}

\begin{figure}[h]
		\centering
		\includegraphics[width=0.9\textwidth]{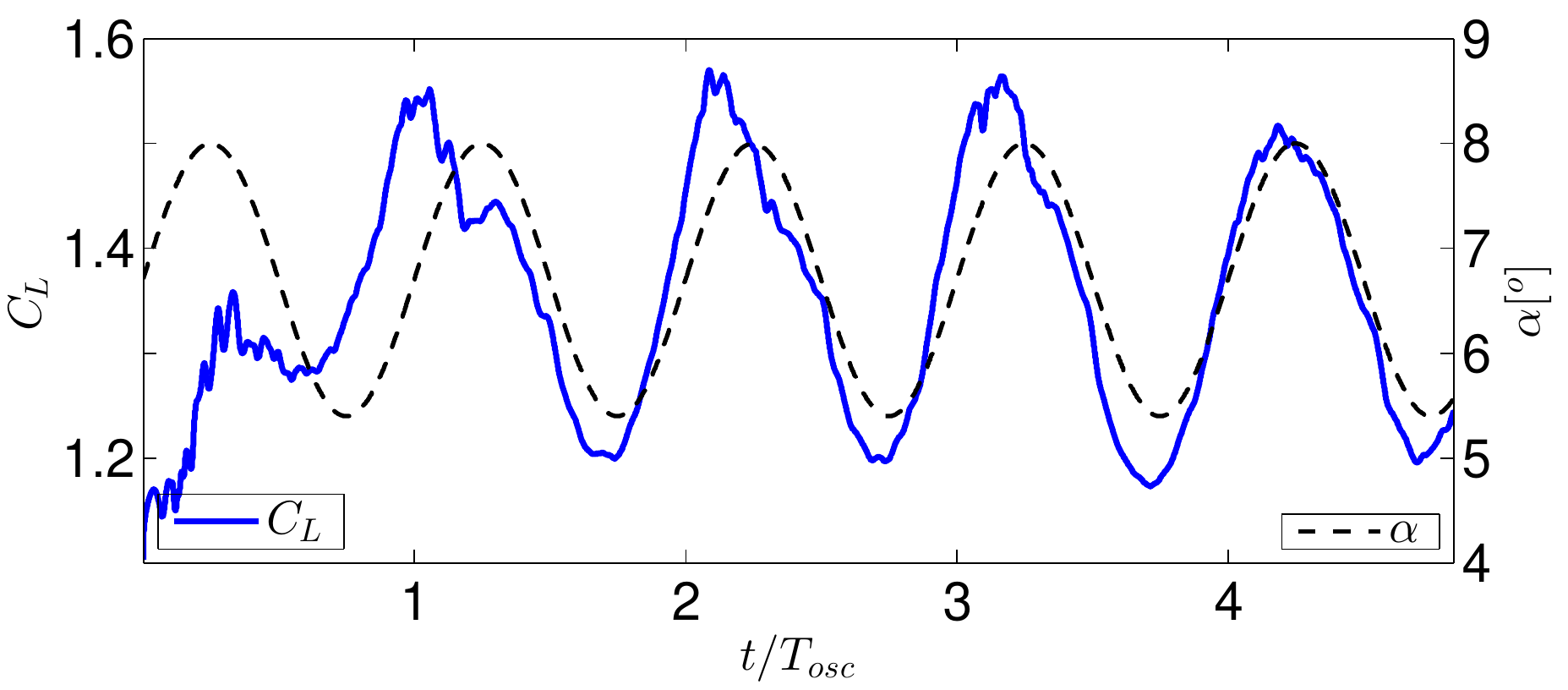}
		\caption{Coefficient of lift ($C_{L} \textcolor{blue}{-}$) and angle of attack $(\alpha \textcolor{black}{--})$ variation with time. $C_{L}$ is on the left axis while $\alpha$ is on the right axis.}
		\label{fig:cl-time-alpha}
\end{figure}

Once the static characteristics of the airfoil are obtained, the dynamic effects on the boundary layer are investigated by pitching the airfoil about a mean angle $\alpha_{0}=6.7^{\circ}$ with an amplitude of $\Delta\alpha=1.3^{\circ}$. The reduced frequency of oscillation is $k=0.5$ and the pitch axis is located at $(x_{0},y_{0})=(0.35,0.034)$. The reduced frequency is defined as $k=\Omega b/U_{0}$, where $\Omega$ is the angular frequency of oscillation, $b$ is the semi-chord length and $U_{0}$ is the free-stream velocity. The motion of the airfoil is prescribed by equation~\ref{eqn:alpha_rule}. The pitching motion corresponds to an oscillation time period of $T_{osc}=2\pi$.
\begin{equation}
	\alpha = \alpha_{0} + \Delta\alpha\sin(\Omega t).
	\label{eqn:alpha_rule}
\end{equation}
The time variation of the coefficient of lift ($C_{L}$) is shown in figure~\ref{fig:cl-time-alpha}. The initial phase of pitching motion is carried out using a lower resolution (polynomial order $N=5$) to simulate the initial transient period of the flow at a lower computational cost. The polynomial order is then smoothly raised to $N=11$ before the fourth pitch cycle. Due to the fairly large separation at the trailing edge, effects of transition movement and turbulence, successive pitch cycles are not expected to have identical behavior, however some of qualitatively repeating trends can be observed. The behavior of the lift coefficient shows a chaotic but qualitatively repeating pattern where $C_{L}$ shows a smooth increase during the pitch-up motion, with strong secondary effects occurring near the maxima of the pitch cycles. Similarly in the pitch-down phase the lift decreases smoothly with secondary effects again becoming important at the minima of the pitch-cycles.
Note that the lift appears to be marginally ahead in phase as compared to the instantaneous angle of attack. This is in fact a feature of pitch oscillations at slightly high reduced frequencies. The linear flutter theory by \cite{theodorsen35} divides the unsteady response of a pitching airfoil into added-mass and circulatory components. While the circulatory component lags the instantaneous angle of attack, the added-mass component exhibits a phase gain (for pure pitch oscillations). At higher reduced frequencies the added-mass components dominate the unsteady response and thus lead to a net phase gain in the lift response. Variation of the theoretical unsteady lift phase with reduced frequencies as well as comparisons with experimental data from \cite{halfman52} and \cite{rainey57} can be found in \cite{leishman00}.

\begin{figure}[h]
	\centering
	\begin{subfigure}[t]{0.49\textwidth}
		\centering
		\includegraphics[width=1\textwidth]{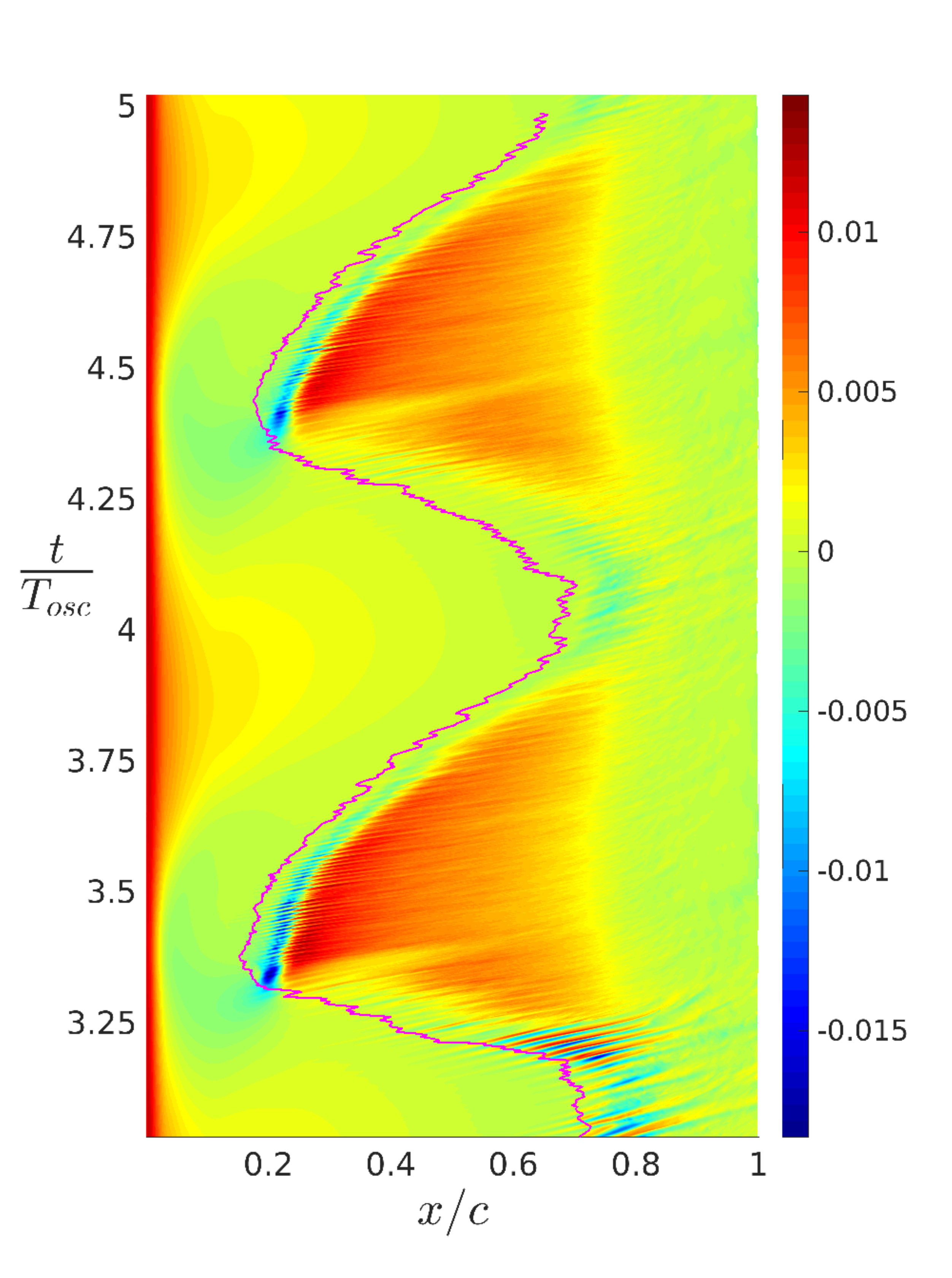}
		\caption{Wall-shear stress ($\tau_{w}$).}. 
		\label{fig:cf-time}
	\end{subfigure}
	\begin{subfigure}[t]{0.49\textwidth}
		\centering
		\includegraphics[width=1.\textwidth]{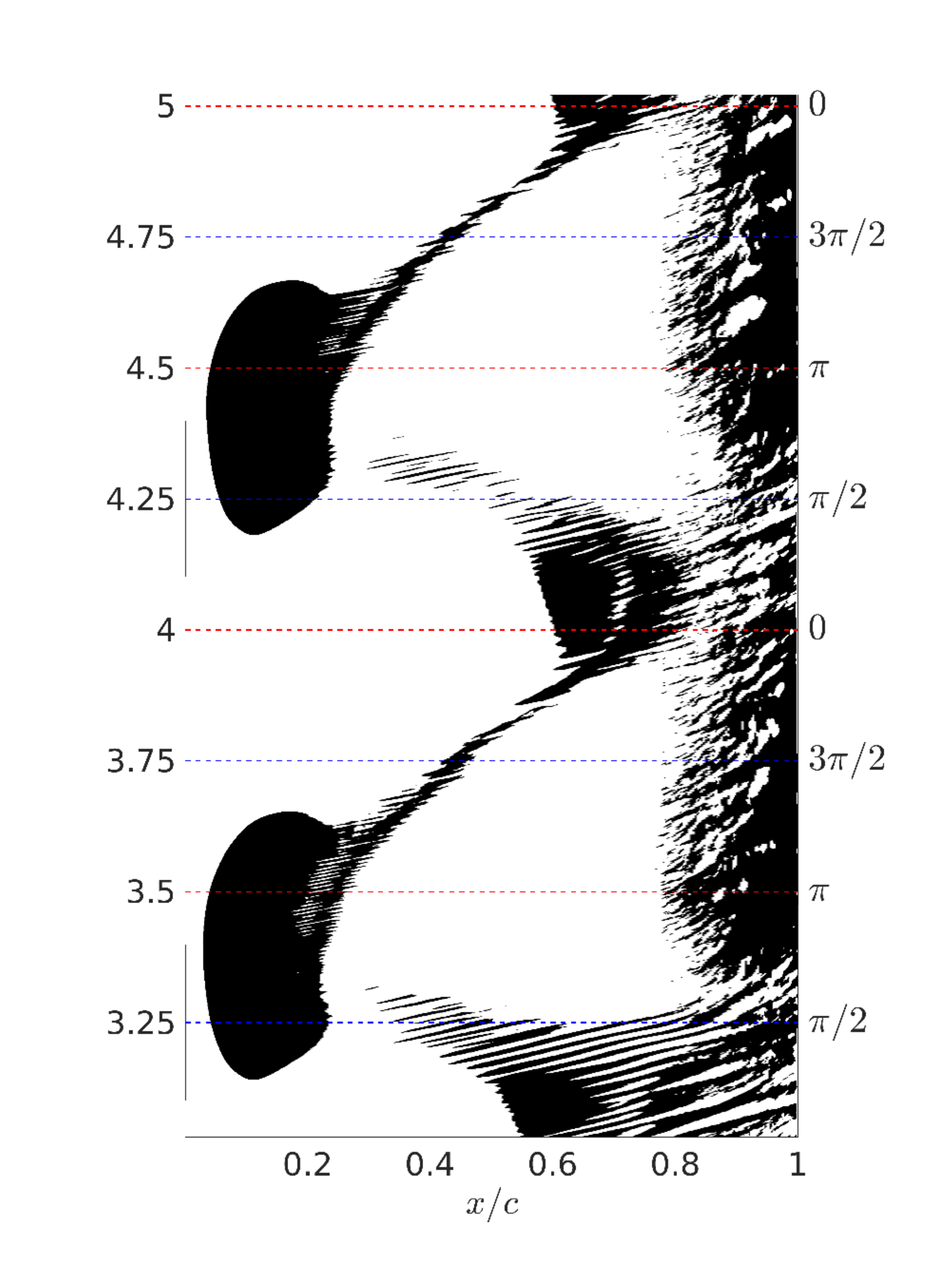}
		\caption{Separated regions (black).} 
		\label{fig:separation-time}
	\end{subfigure}
	\caption{(a) Space-time plot for the wall-shear values ($\tau_{w}$) and (b) separated flow regions. The values are obtained from the instantaneous flow averaged over the spanwise direction. Magenta line in (a) denotes the calculated transition location. Horizontal blue dashed lines in (b) represent the extrema of the angle of attack, while the red dashed lines represent phases corresponding to mean angle of attack.}
	\label{fig:space-time}
\end{figure}

In order to understand the time variation of the spatially developing boundary layer on the airfoil, we analyze the space-time evolution of the instantaneous spanwise averaged wall-shear stress. The space-time surface plot is shown in figure~\ref{fig:cf-time}, which spans the fourth and fifth pitch cycles. The color specifies the value of wall-shear stress on the suction side of the airfoil. Regions with color intensity strongly towards red are indicative of high shear and thus turbulent flow. The exception to the rule being the region close to the leading edge where the flow is laminar but a high shear region exists due to the extremely thin boundary layer close to the stagnation point. The same space-time surface is shown again as a binary colored surface plot in figure~\ref{fig:separation-time}, where black colored regions indicate negative wall-shear stress and hence separated flow, while the white region corresponds to locations with attached flow ($\tau_{w}>0$).

It is obvious from the two plots in figure~\ref{fig:space-time}, that the developing boundary layer on the airfoil exhibits a dynamically rich response to small-amplitude pitch oscillations, with different key boundary layer characteristics controlling the dynamics of the flow in different phases of the pitch cycle. We identify some of the key boundary-layer characteristics to paint an over-all picture of the dynamics. A persistent trailing-edge separation can be identified in figure~\ref{fig:separation-time} beyond $x/c>0.8$. The trailing-edge separation does not exhibit reverse flow $100\%$ of the time, as can be seen from the white patches dispersed between largely black colored regions. An isolated separated region (distinct from the trailing edge separation) is observed at $x/c\approx0.6$ at times $t/T_{osc}\approx3$ and $t/T_{osc}\approx4$. This is identified as a trailing-edge LSB. This LSB is short lived in time, existing for slightly less than a quarter of the pitch-cycle. A large separated region near the leading edge is a leading-edge LSB, which persists much longer in time, spanning nearly half a pitch cycle. Evident from figure~\ref{fig:cf-time} is that the transition point changes substantially throughout the pitch cycle. Interestingly, the flow over the airfoil differs significantly during the pitch-up and the pitch-down phases for the same angle of attack. For example, when the instantaneous angle of attack is at phase $\phi=0\ (t/T_{osc}=3,\ 4)$, which represents mean angle of attack but in the pitch-up phase, the flow over the airfoil is mostly laminar up to $x/c\approx0.7$. On the other hand, for a phase of $\phi=\pi\ (t/T_{osc}=3.5,\ 4.5)$, representing the airfoil at the mean angle of attack but in the pitch-down phase, the flow is almost entirely turbulent with the start of the turbulent region approximately at $x/c\approx0.22$.

\subsection{Transition location}

\begin{figure}[h]
	\centering
	\begin{subfigure}[t]{0.49\textwidth}
		\centering
		\caption{}. 
		\includegraphics[width=1\textwidth]{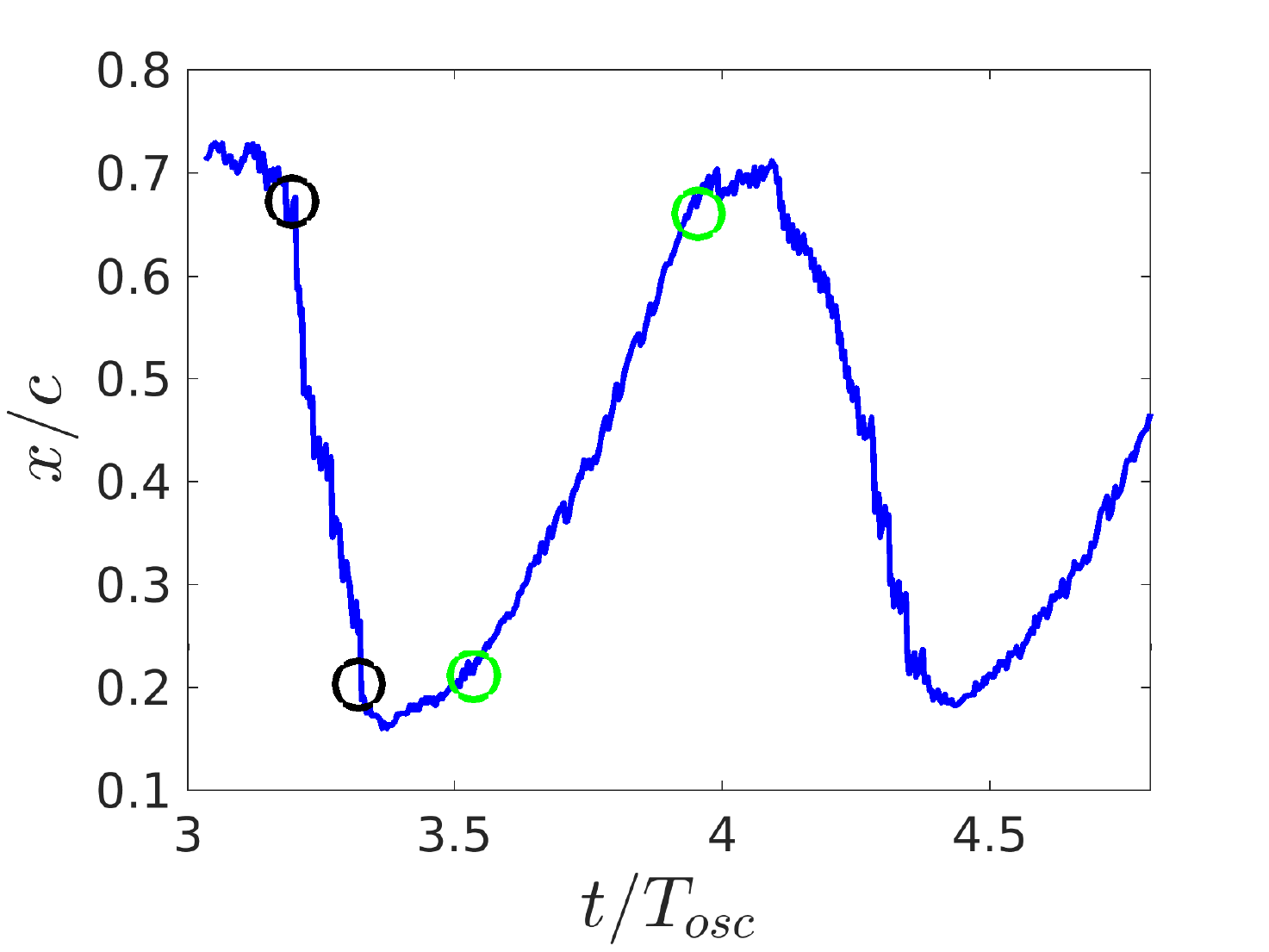}
		\label{fig:transition_time}
	\end{subfigure}
	\begin{subfigure}[t]{0.49\textwidth}
		\centering
		\caption{} 		
		\includegraphics[width=1\textwidth]{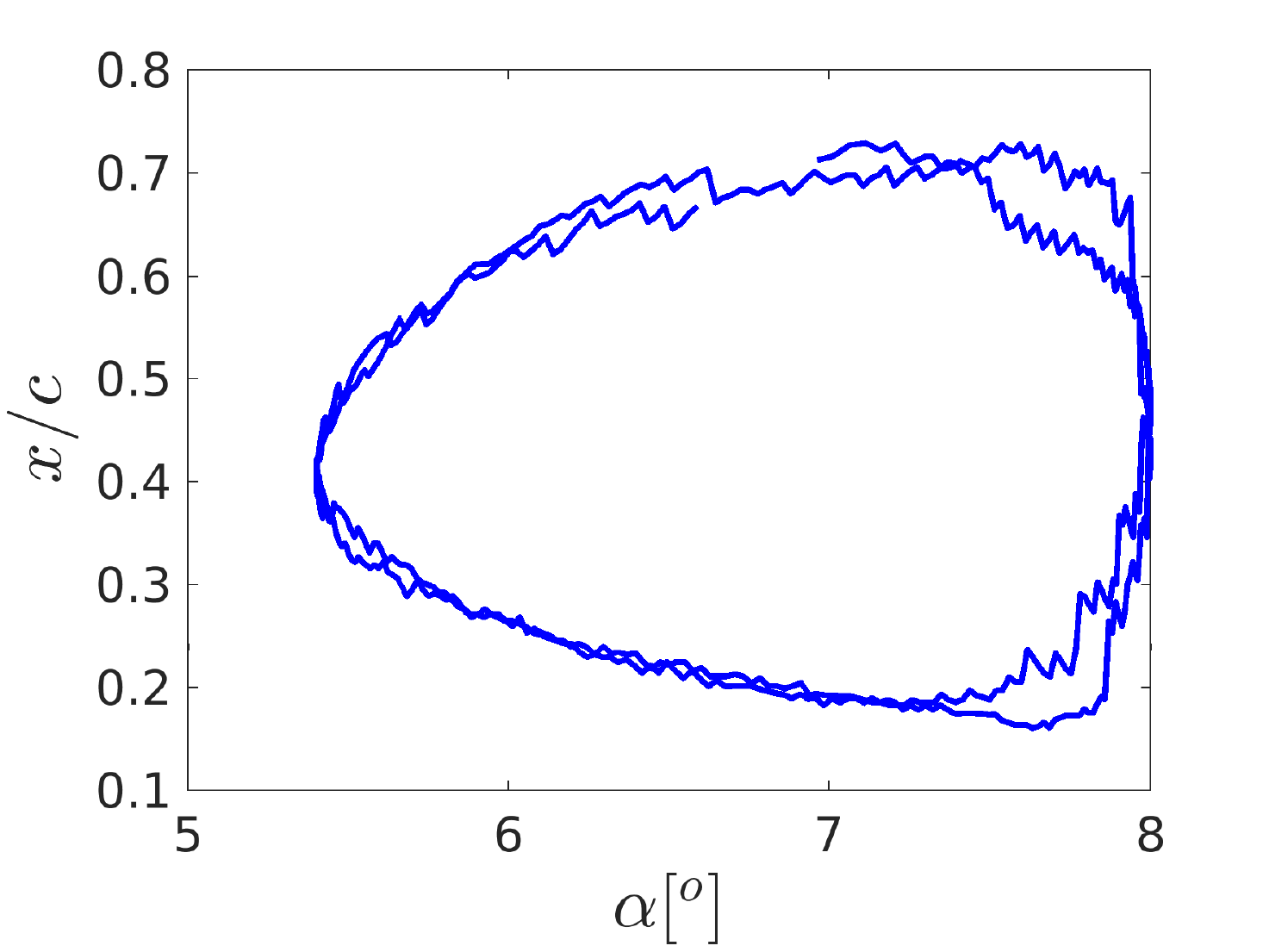}
		\label{fig:transition_alpha}
	\end{subfigure}
	\caption{Time variation (a) and phase portrait with $\alpha$ (b) of transition location evaluated using the empirical criteria for $|\overline{u'v'}|$. The circles mark the points used to approximate the upstream (black circles) and downstream (green circles) velocities of the transition point.}
	\label{fig:transition}
\end{figure}

\begin{figure}
	\centering
	\begin{subfigure}[t]{0.49\textwidth}
		\begin{overpic}[width=1\textwidth]{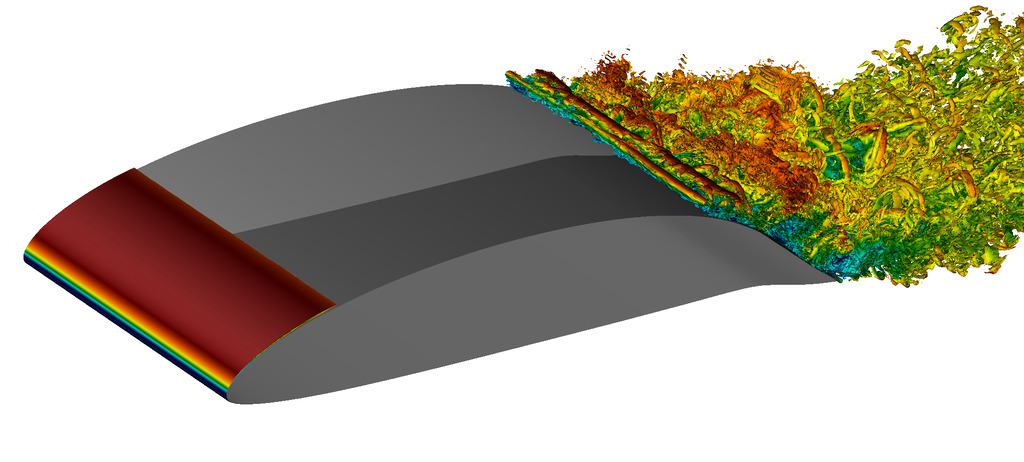}
			\put(60,42){\includegraphics[width=0.45\textwidth,height=0.20\textwidth]{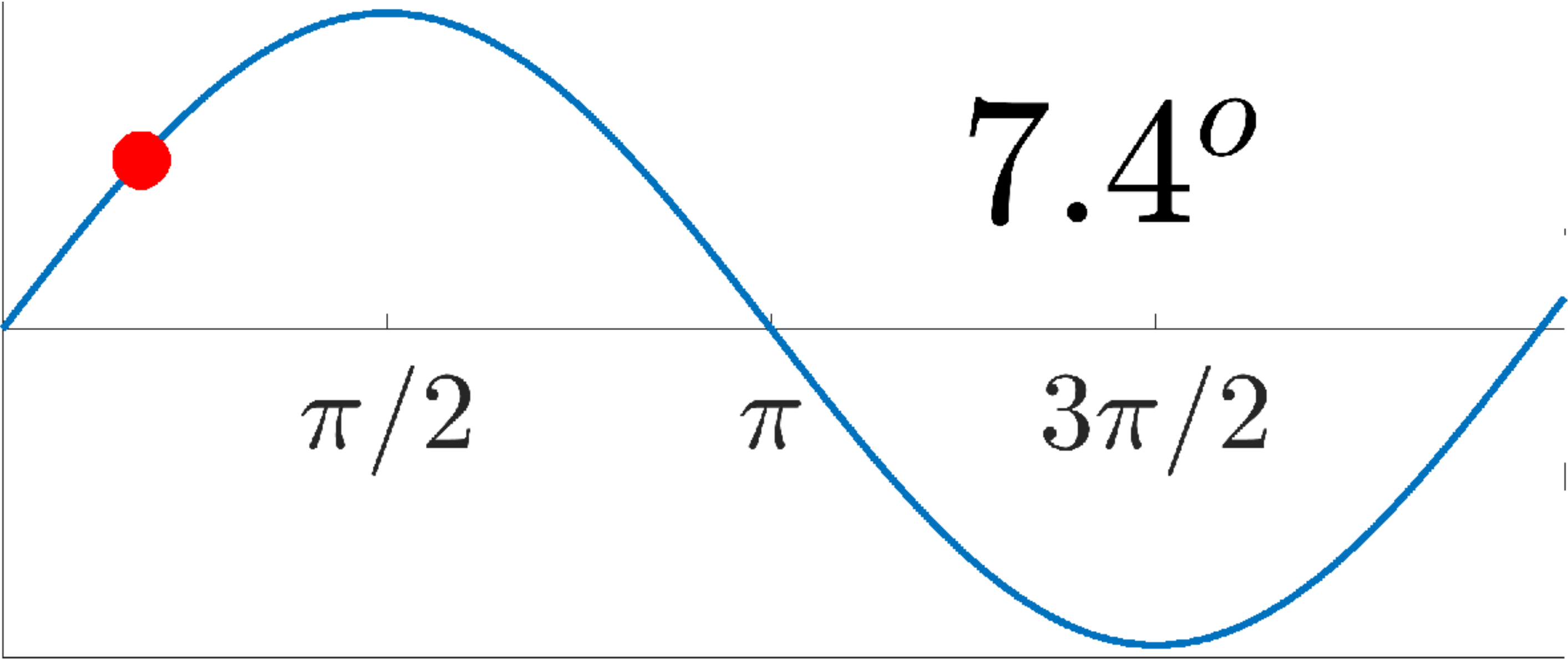}}
		\end{overpic}		
		\label{fig:pitchup_1}
	\end{subfigure}
	\begin{subfigure}[t]{0.49\textwidth}
		\centering
		\begin{overpic}[width=1\textwidth]{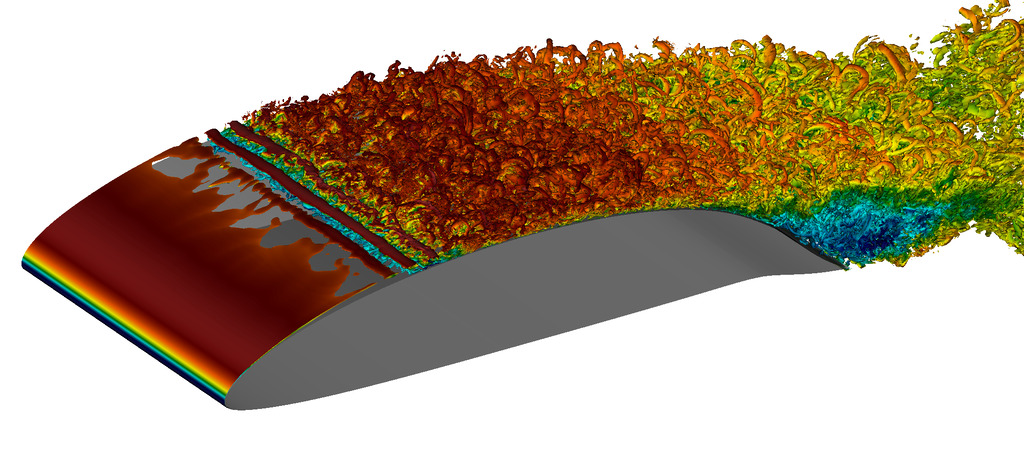}
			\put(60,42){\includegraphics[width=0.45\textwidth,height=0.20\textwidth]{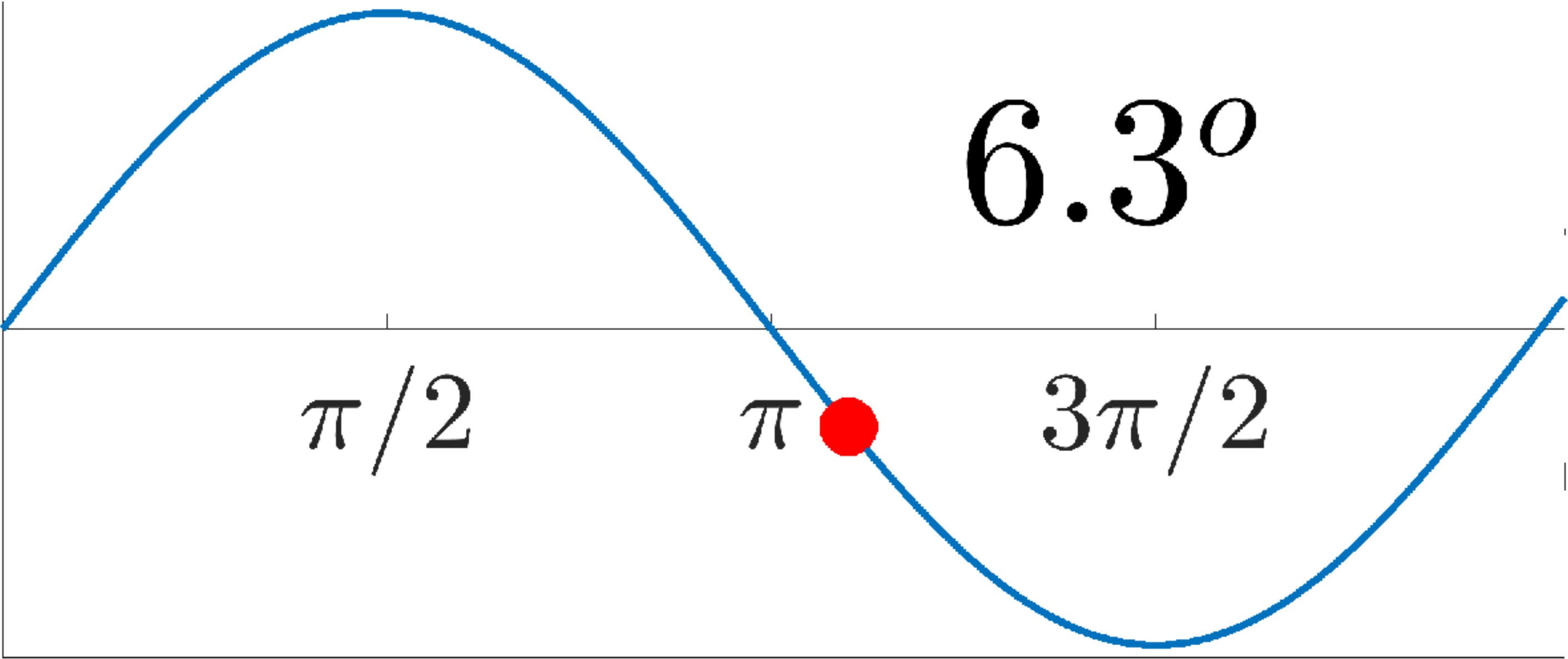}}
		\end{overpic}
		\label{fig:relaminar_1}
	\end{subfigure}
	\begin{subfigure}[t]{0.49\textwidth}
		\centering
		\begin{overpic}[width=1\textwidth]{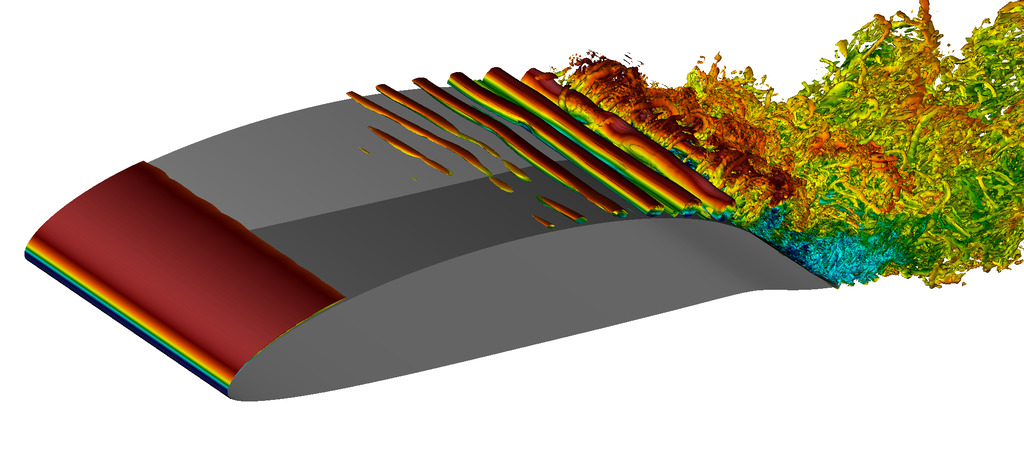}
			\put(60,42){\includegraphics[width=0.45\textwidth,height=0.20\textwidth]{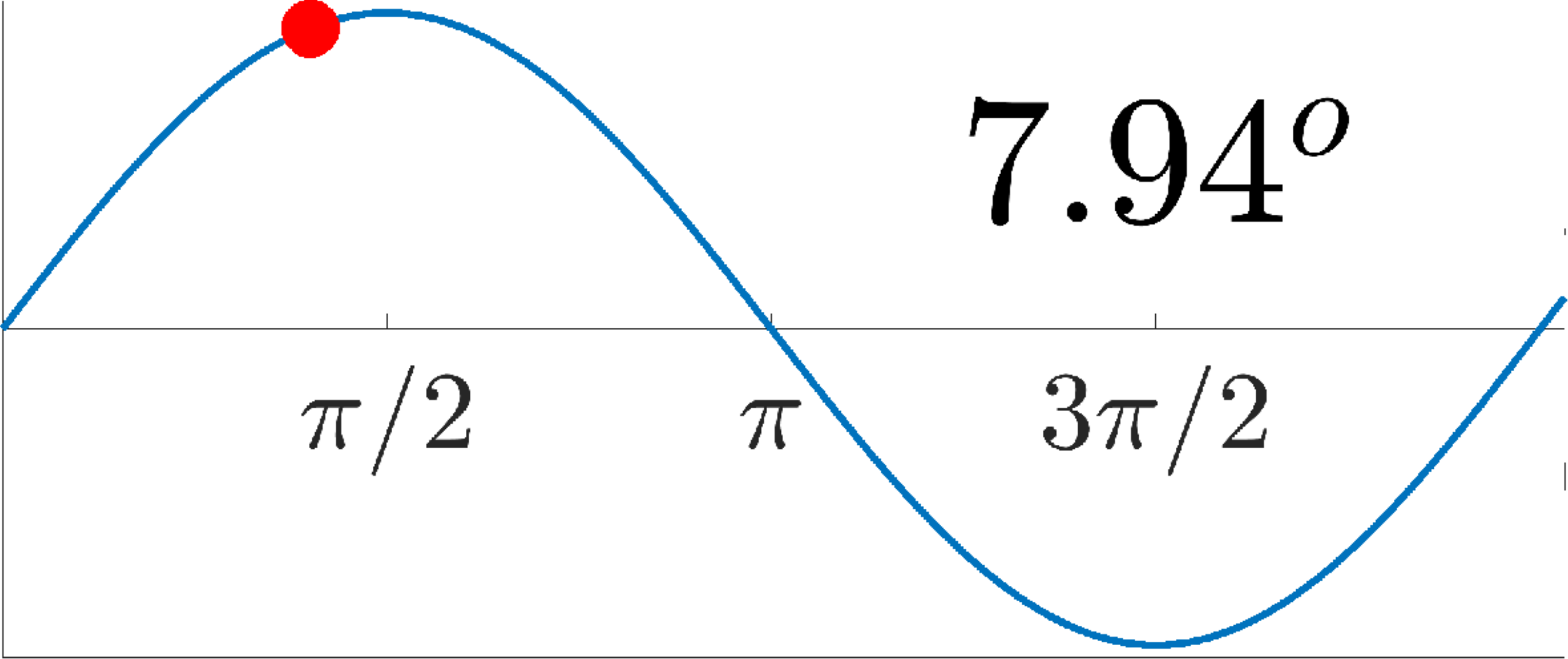}}
		\end{overpic}		
		\label{fig:pitchup_2}
	\end{subfigure}
	\begin{subfigure}[t]{0.49\textwidth}
		\centering
		\begin{overpic}[width=1\textwidth]{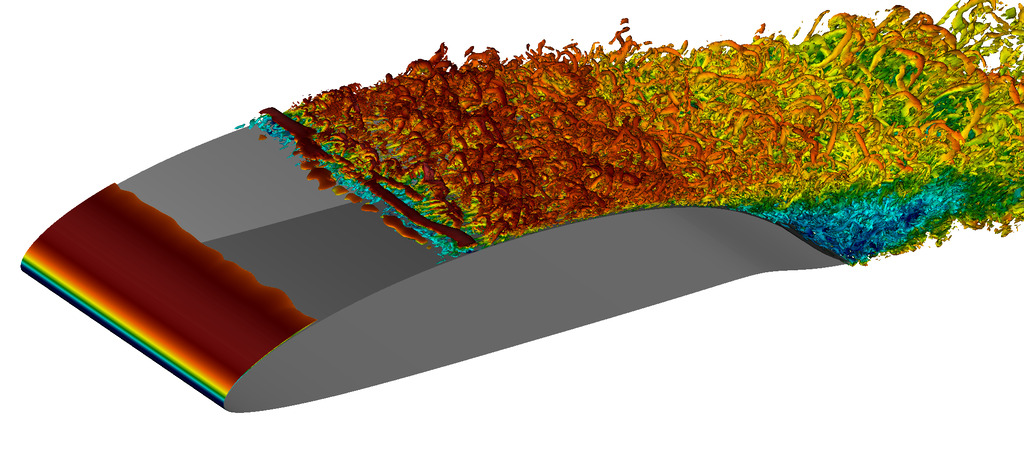}
			\put(60,42){\includegraphics[width=0.45\textwidth,height=0.20\textwidth]{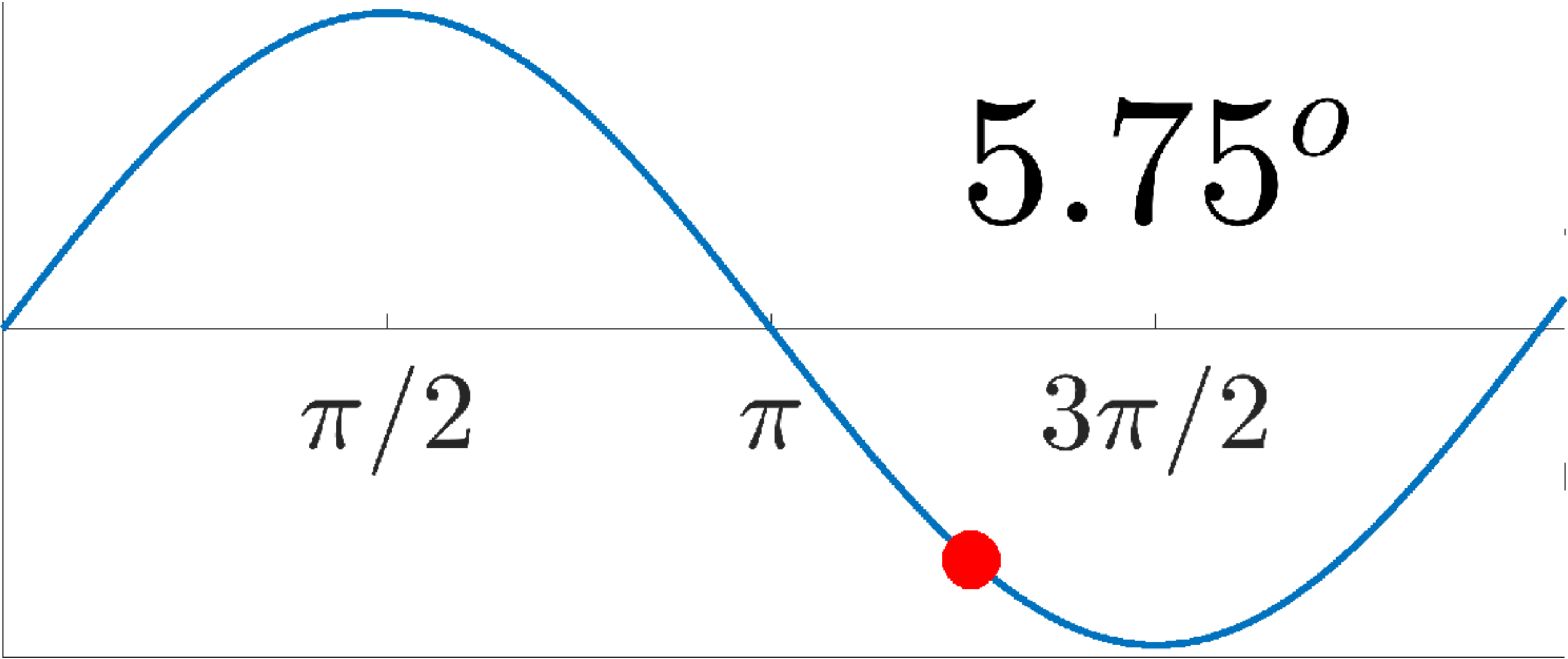}}
		\end{overpic}		
		\label{fig:relaminar_2}
	\end{subfigure}
	\begin{subfigure}[t]{0.49\textwidth}
		\centering
		\begin{overpic}[width=1\textwidth]{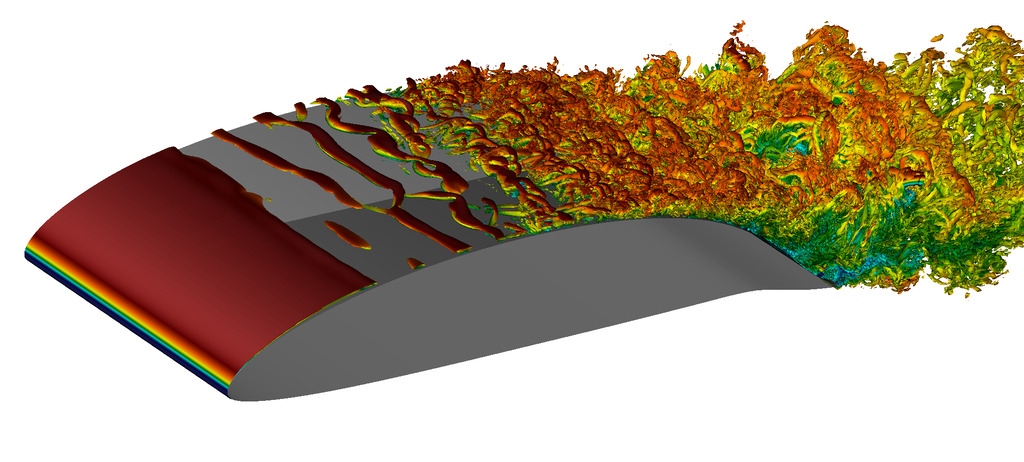}
			\put(60,42){\includegraphics[width=0.45\textwidth,height=0.20\textwidth]{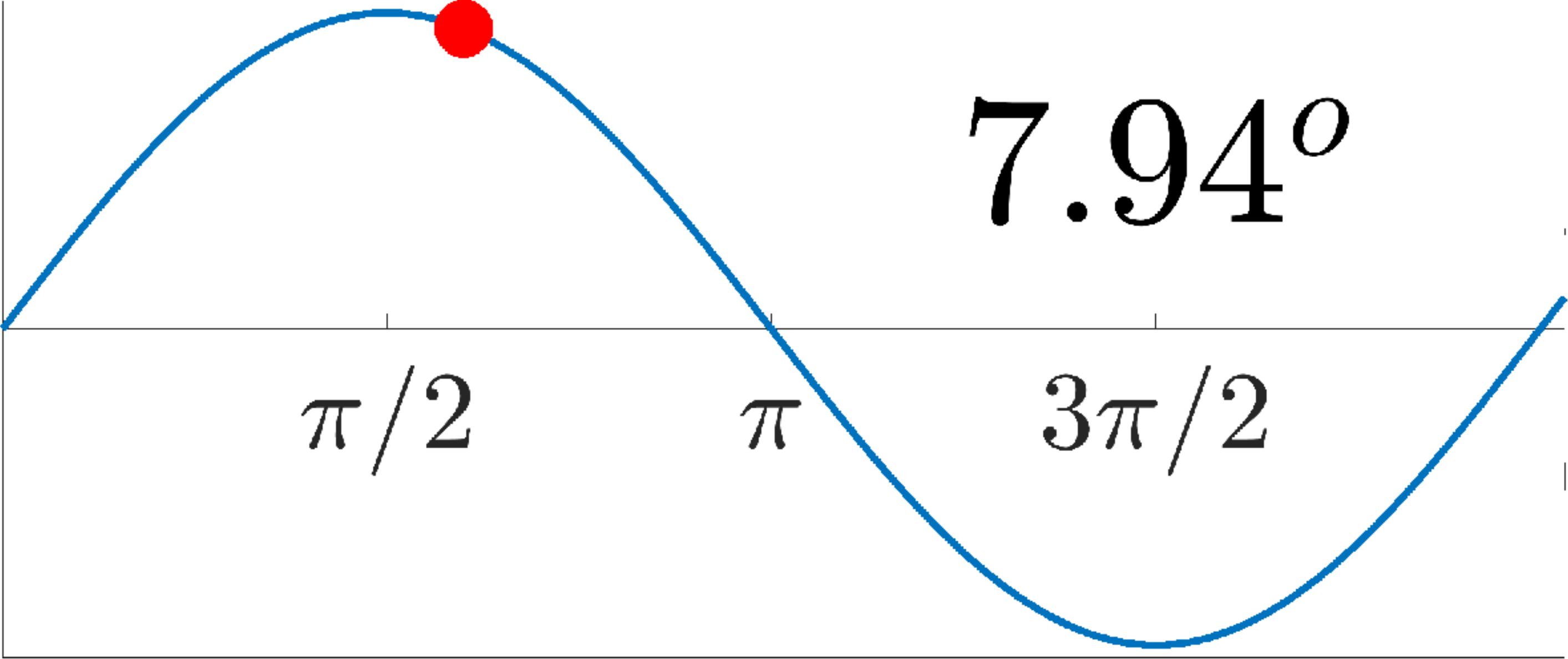}}
		\end{overpic}
		\label{fig:pitchup_3}
	\end{subfigure}
	\begin{subfigure}[t]{0.49\textwidth}
		\centering
		\begin{overpic}[width=1\textwidth]{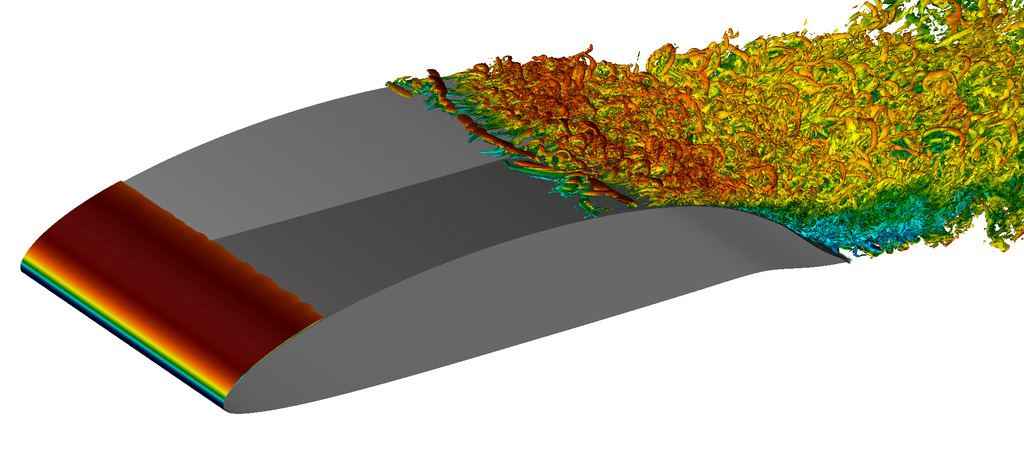}
			\put(60,42){\includegraphics[width=0.45\textwidth,height=0.20\textwidth]{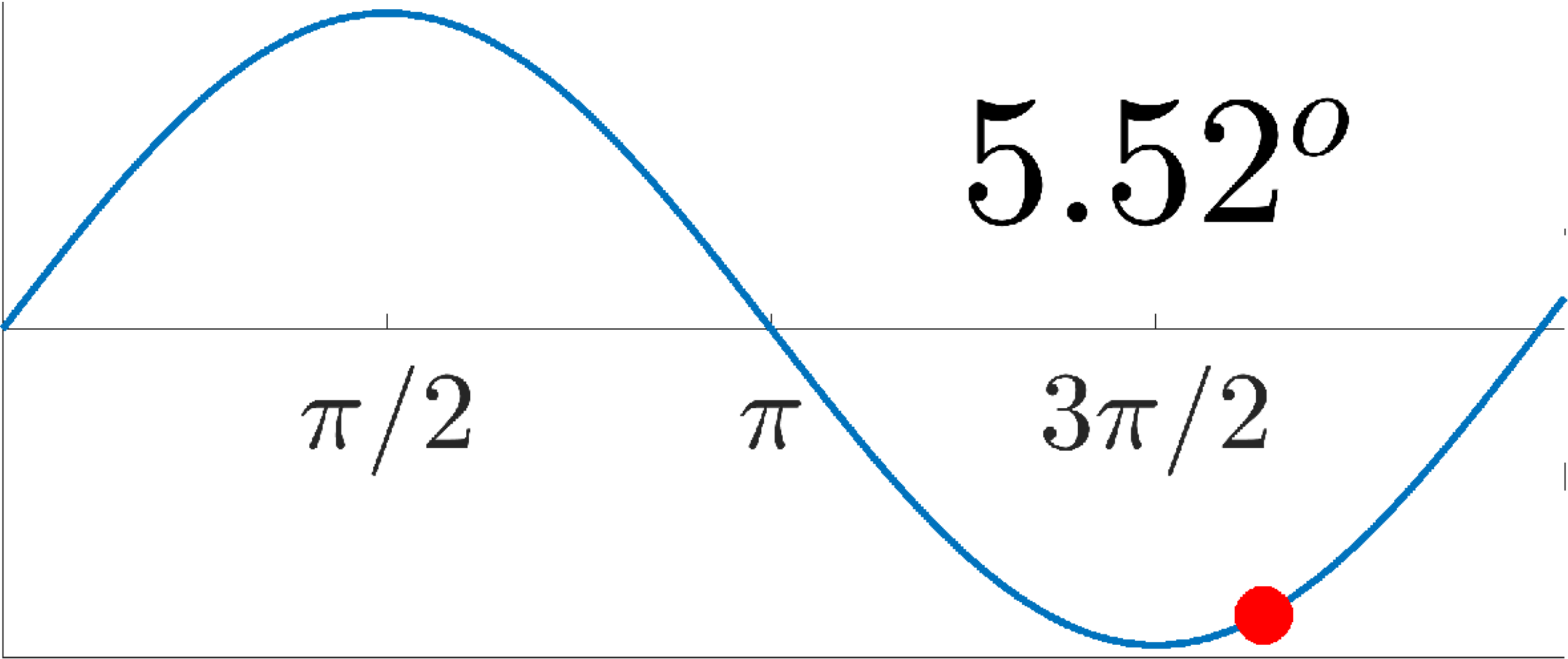}}
		\end{overpic}
		\label{fig:relaminar_3}
	\end{subfigure}
	\begin{subfigure}[t]{0.49\textwidth}
		\centering
		\begin{overpic}[width=1\textwidth]{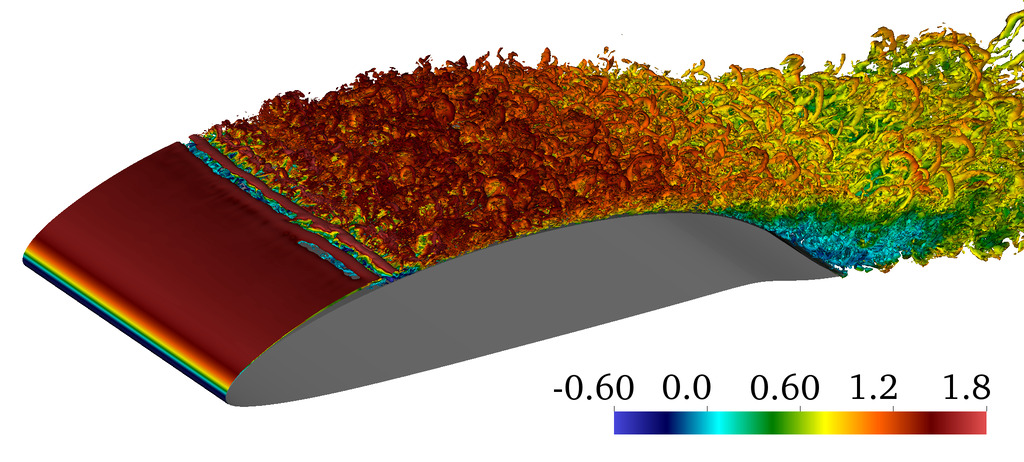}
			\put(60,42){\includegraphics[width=0.45\textwidth,height=0.20\textwidth]{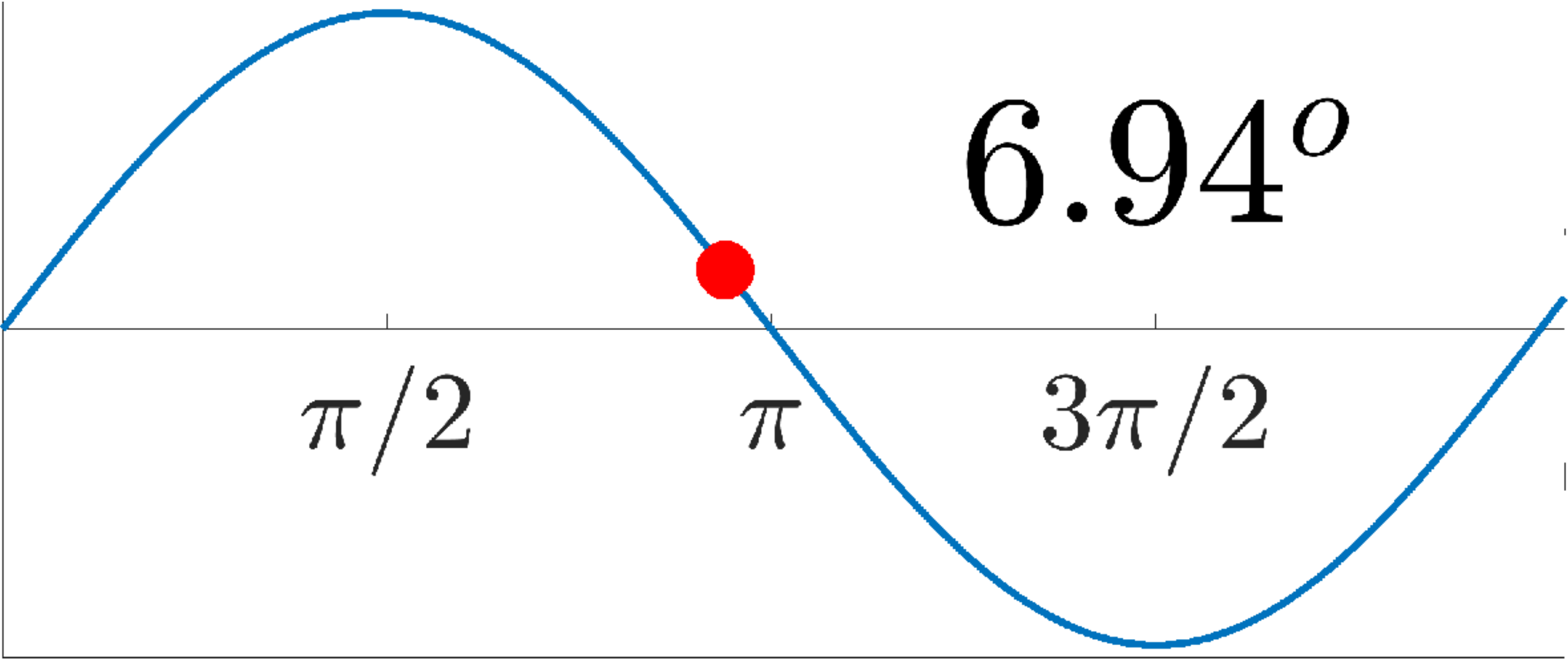}}
		\end{overpic}
		\label{fig:pitchup_4}
	\end{subfigure}
	\begin{subfigure}[t]{0.49\textwidth}
		\centering
		\begin{overpic}[width=1\textwidth]{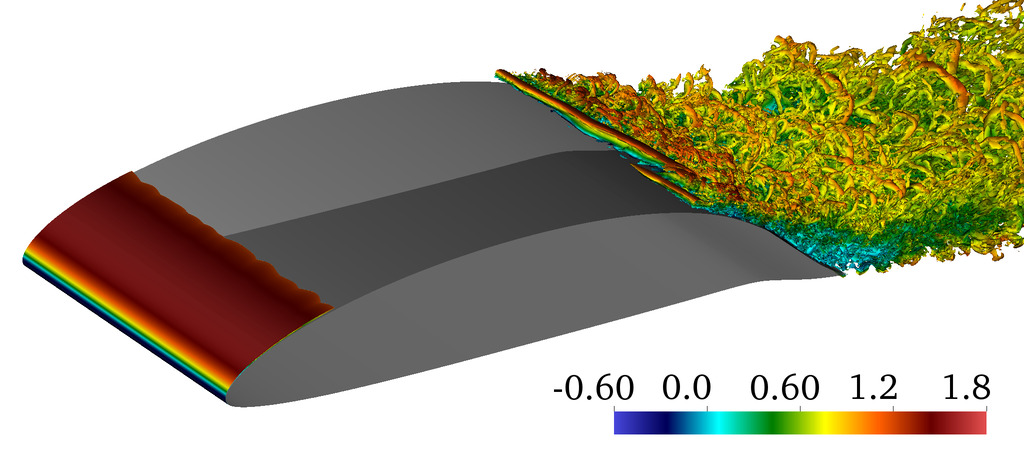}
			\put(60,42){\includegraphics[width=0.45\textwidth,height=0.20\textwidth]{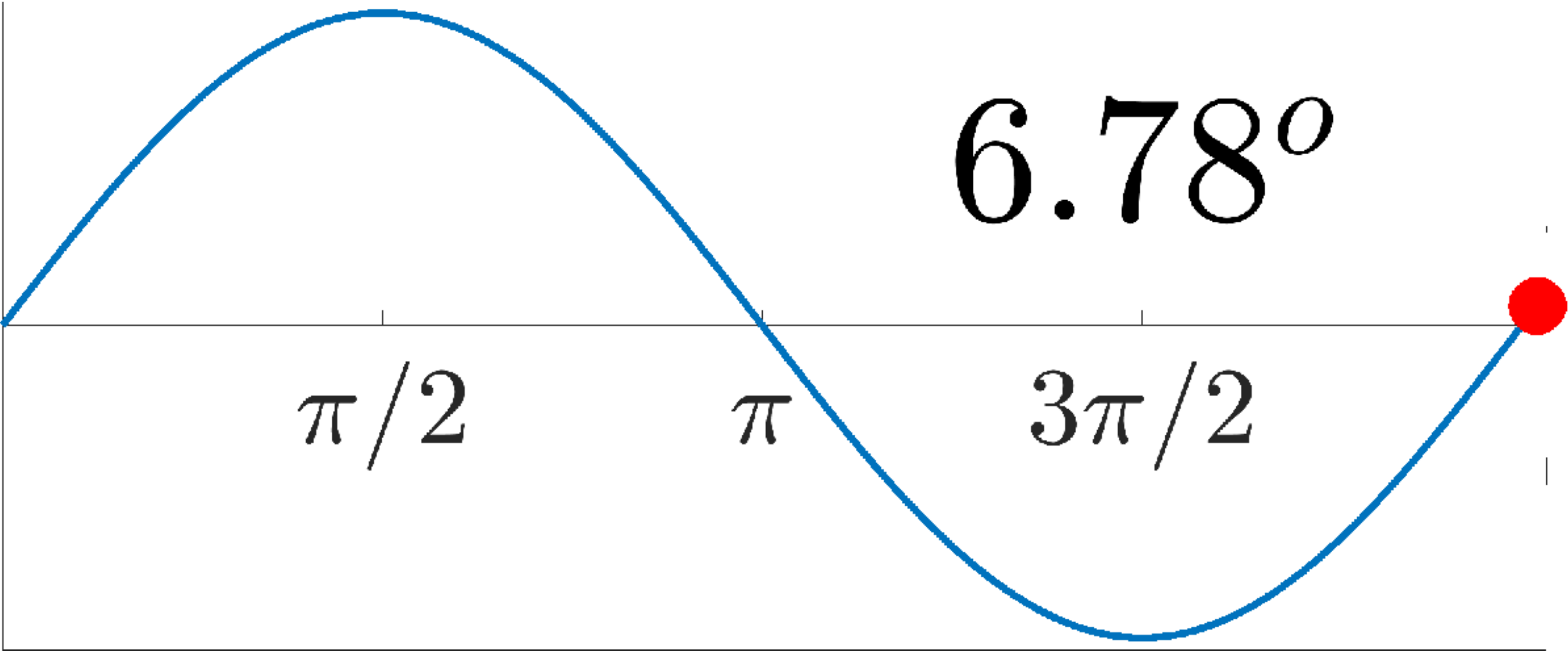}}
		\end{overpic}
		\label{fig:relaminar_4}
	\end{subfigure}	
	\caption{Visualization of instantaneous $\lambda_{2}$ structures at different phases of the pitch cycle. The figures on the left are during the phase of the pitch cycle when the transition is moving upstream. On the right the instantaneous snapshot correspond to the re-laminarization phase as transition moves downstream. The instantaneous angle of attack and the phase of oscillation is given by the small inset on top of each panel.}	
	\label{fig:transition_la2}
\end{figure}

In the present work we focus on the variation of transition location throughout the pitch cycles. Since the flow case is unsteady, and the transition location does not remain fixed, a criteria based on the instantaneous state of the flow is needed to determine the transition location. The details of the evaluation of the transition location can be found in \ref{app:A}. The temporal variation and the phase portrait of the calculated transition location are shown in figure~\ref{fig:transition}. The magenta line in figure~\ref{fig:space-time} shows the calculated transition locations superposed on the space-time plot of the wall-shear stress. The empirical transition locations are consistent with the picture of wall-shear stress, with transition marginally preceding regions of turbulent flow.
The superposed plots in figure~\ref{fig:space-time} indicate that the two LSBs play a dominating role in the flow dynamics and that transition location is governed by the characteristics of these LSBs. Figure~\ref{fig:transition_la2} (left) shows the iso-contours of instantaneous vortical structures, identified by the $\lambda_{2}$ method \citep{jeong95}, at four different times during the pitch-up cycle when the transition is moving upstream. This phase is marked by the appearance of a leading-edge LSB which grows in size. The top figure shows the flow state near the mean angle of attack ($t/T_{osc}=3.09$) during the pitch-up phase. The flow is mostly laminar across the airfoil with no structures observed prior to flow transition (close to trailing edge) and there is no leading-edge LSB. The high adverse pressure gradient near the trailing edge causes the laminar flow to easily separate, forming a LSB and flow transitions over this separated shear layer. Figure~\ref{fig:transition_la2} (left) from top to bottom shows the part of the oscillation cycle when transition is moving upstream at time instants of $t/T_{osc}=3.09,\ 3.2,\ 3.3$ and $3.47$, which corresponds to instantaneous angles of $\alpha=7.4^{o},7.94^{o},7.94^{o}$ and $6.94^{o}$ respectively. Note that pitch-up phase completes at $t/T_{osc}=3.25$ when the airfoil is at the highest angle of attack. Thus upstream movement of transition starts nearly at the end of the pitch-up phase and continues to move upstream even during the pitch-down cycle. The laminar separation bubble close to the trailing edge ceases to exist as transition moves upstream. At $t/T_{osc}=3.47$ the flow transition is seen to occur on the separated shear layer of the leading-edge LSB (bottom left in figure~\ref{fig:transition_la2}).

As the airfoil progresses through the pitch-down cycle the leading-edge LSB shrinks in size and the transition point then starts moving downstream, initiating the process of flow re-laminarization (figure~\ref{fig:transition_la2} right, top to bottom). The leading-edge LSB eventually disappears, although the transition point moves downstream of the LSB before it completely disappears. The flow over the airfoil is not completely re-laminarized even when the airfoil is at the lowest angle of attack and the re-laminarization process continues into the pitch-up phase. There is a marked asymmetry between the upstream and downstream movement of the transition point. An approximate velocity for both the upstream and downstream motion of the transition point is calculated across the points marked by circles in figure~\ref{fig:transition_time} which correspond to transition movement with near constant velocity. The velocity of upstream transition movement is calculated across the black circles and is equal to $V^{tr}_{u}=-0.60$ while the velocity of the downstream motion of transition is calculated across the green circles and is equal to $V^{tr}_{d}=0.17$. Thus the upstream spread of turbulent regions is much faster than flow re-laminarization. 

\begin{figure}
	\centering
	\begin{subfigure}[t]{0.48\textwidth}
		\centering
		\includegraphics[width=1\textwidth]{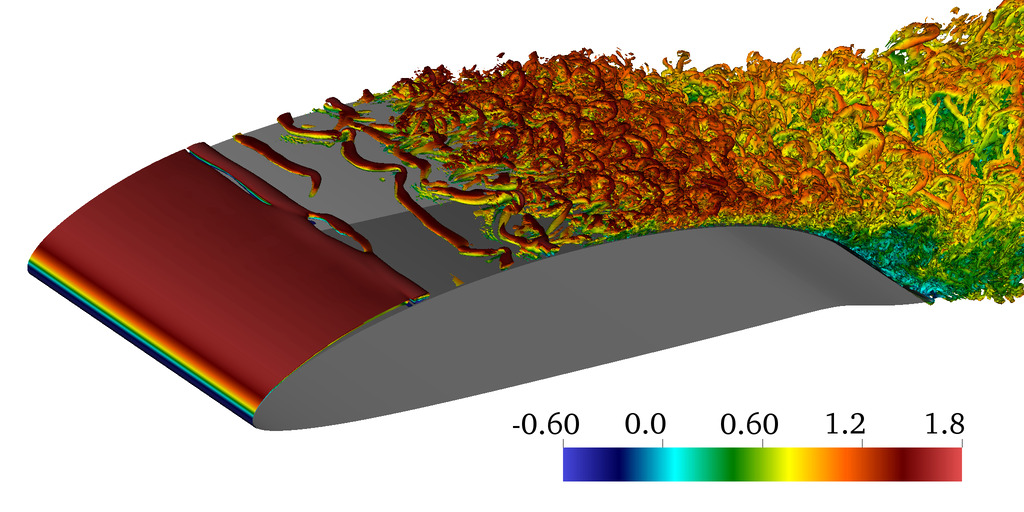}
		\label{fig:upstream_1}
	\end{subfigure}
	\begin{subfigure}[t]{0.48\textwidth}
		\centering
		\includegraphics[width=1\textwidth]{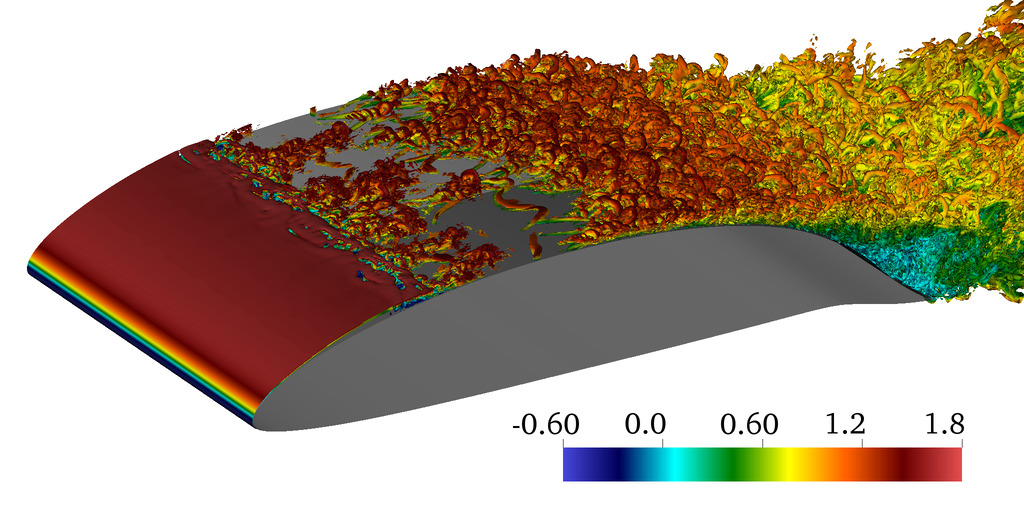}
		\label{fig:upstream_2}
	\end{subfigure}
	\caption{Comparison of boundary layer transition at two different time instants of $t/T_{osc}=3.33$ (left) and $t/T_{osc}=3.35$ (right).}
	\label{fig:transition_complex}
\end{figure}

Specifying a velocity of transition movement implies that the transition location changes smoothly with time. This is true for the downstream movement of transition, however the final stages of the upstream movement appear to be more complex. Figure~\ref{fig:transition_complex} shows the instantaneous vortical structures at two time instants during the upstream movement phase. In the left figure ($t/T_{osc}=3.33$) a single connected region of turbulence can be observed which is preceded by a laminar region identified by the near absence of small vortical structures. This region starts at about $40\%$ and the entire boundary layer downstream is turbulent. On the right figure ($t/T_{osc}=3.35$) there is a similar large connected region of turbulence, spreading from approximately $40\%$ of the chord. However there is also a nascent region of turbulence starting at $x/c\approx0.2$. These two regions are separated by a patch of laminar flow. The figure on the right then has two distinct locations where transition to turbulence takes place. After a short while, turbulence spreads downstream from this newly formed transition location and eventually the entire boundary layer downstream is turbulent. This flow state, where two distinct turbulent regions can be identified exists only for a short duration and by $t/T_{osc}=3.4$ no intermediate regions of laminar flow can be identified. However the emergence of two distinct transition locations indicates that at least two competing mechanisms exist for the growth of disturbances in the boundary layer and that their relative strength changes during the pitch cycle. Given that this new transition occurs at the separated shear layer of the leading-edge LSB, the stability properties of the LSB are likely to play a role in the emergence of this new transition point. 

\subsection{Stability characteristics of the laminar separation bubble}

Stability characteristics of steady laminar separation bubbles have been studied with different perspectives by several previous authors like \cite{hammond98,alam00,marxen03,lang04,marxen10,jones08,jones10,marxen12,marxen13} \textit{etc.} In unsteady cases, the stability characteristics of the leading-edge LSB can help shed some light on the changing transition locations throughout the pitch cycle and in particular on the existence of competing mechanisms for transition. The competition between convective and absolute instability characteristics may provide a possible explanation for the transient existence of two distinct points of transition. The change of flow state from laminar to turbulent is often governed by the linear amplification of small disturbances within the boundary layer. For flows with streamwise and spanwise homogeneity the evolution of small perturbations within the boundary layer is governed by the Orr-Sommerfeld equation
\begin{align}
	\bigg[\big(\frac{\partial}{\partial t} + U\frac{\partial}{\partial x}\big)\nabla^{2} - U''\frac{\partial}{\partial x} - \frac{1}{Re}\nabla^{4}	\bigg]v = 0,
\end{align}
where $v$ is the wall-normal component of the velocity fluctuations and $U''$ is the second derivative in the wall-normal direction of the streamwise velocity $U$. Due to Squire's theorem, analyzing the two-dimensional perturbations is sufficient to study the modal stability characteristics. Following \cite{schmid01} and assuming an asantz function for the wall-normal perturbations as
\begin{align}
	v = \tilde{v}(y)e^{i(k_{x}x - \lambda t)},
\end{align}	
results in a dispersion relation between the complex frequency $\lambda$ and the streamwise wavenumber $k_{x}$ given by
\begin{align}
	\bigg[(-i\lambda + ik_{x}U)(\mathcal{D}^{2} - k_{x}^{2}) - ik_{x}U'' - \frac{1}{Re}(\mathcal{D}^{2} - k_{x}^2)^{2}\bigg]\tilde{v} = 0.
	\label{eqn:orr-somerfeld}
\end{align}
Here $\mathcal{D}$ represents the derivative in the wall-normal ($y$) direction. While strictly valid for homogeneous flows, the Orr-Sommerfeld equation has often been used for flows that exhibit weak inhomogeneity such as the Blasius boundary layer and also for the case of laminar separation bubbles \citep{alam00,hammond98,haggmark01b}; \cite{boutilier12,marxen12}.

A temporal stability analysis using a real spatial wavenumber $k_{x}$ results in an eigenvalue problem for the frequency $\lambda$. Resulting complex frequencies with a positive imaginary part signify that the boundary layer is unstable and that small perturbations within the boundary layer would grow in time and cause transition. 
\begin{figure}[h]
	\centering
	\begin{subfigure}[t]{0.48\textwidth}
		\centering
		\includegraphics[width=1\textwidth]{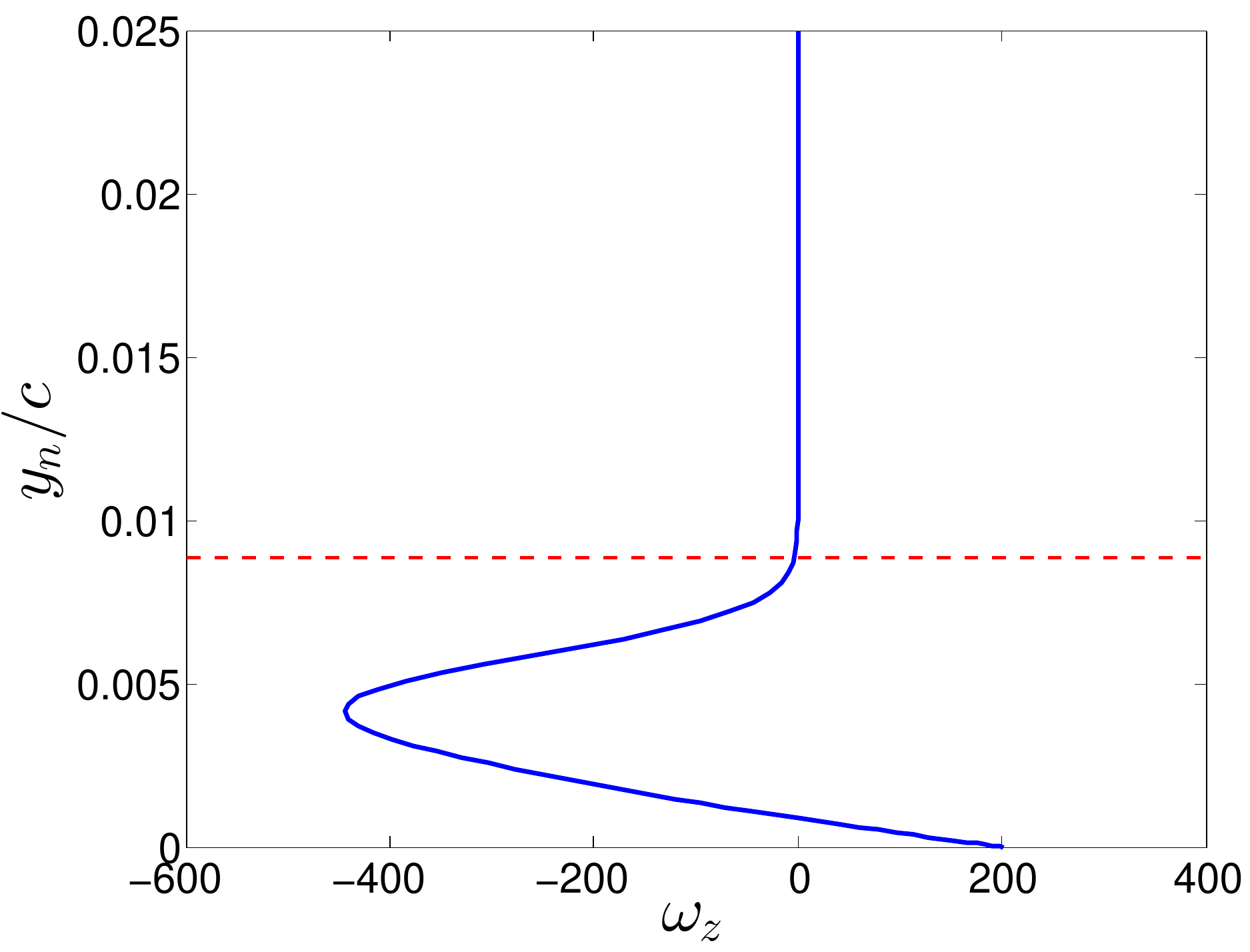}
	\end{subfigure}
	\begin{subfigure}[t]{0.48\textwidth}
		\centering
		\includegraphics[width=1\textwidth]{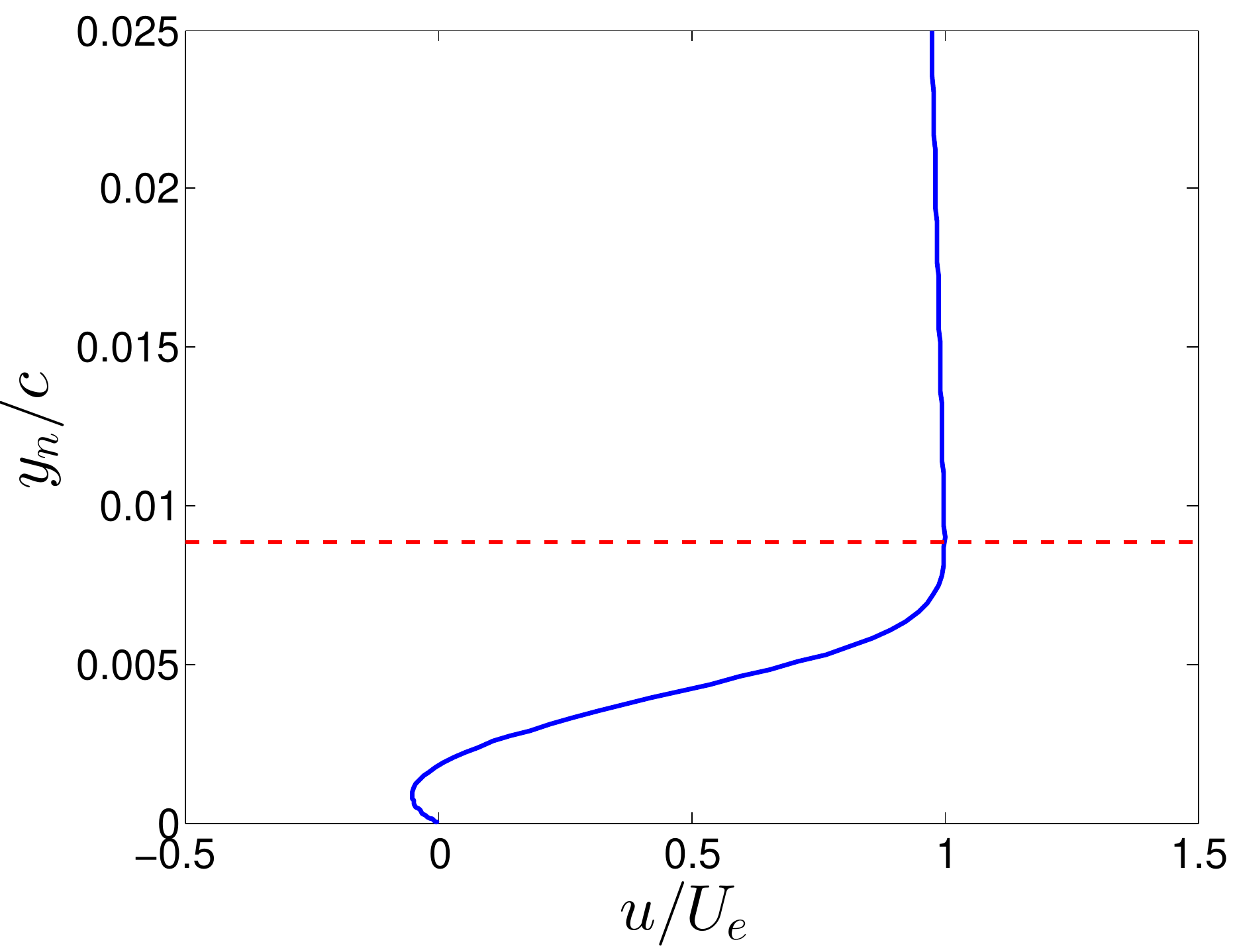}
	\end{subfigure}
	\caption{Wall-normal profiles of vorticity (left) and tangential velocity (right) observed in the leading-edge LSB at $t/T_{osc}=3.25$. Dashed lines mark the boundary layer edge. $y_{n}$ is the distance along the local wall-normal direction.}
	\label{fig:bubble_profiles}
\end{figure}

To explore the time varying stability properties of the LSB, temporal stability analysis of the Orr-Sommerfeld equations is performed using the instantaneous wall-normal profiles of tangential velocity, calculated as per equation~\ref{eqn:inst_q}. Several velocity profiles can be considered for local analysis. \cite{alam00} and \cite{hammond98} in their local analysis associate the stability characteristics of the LSB with the maximum reverse flow intensities. In accordance with the previous studies, we focus on the wall-normal profiles of tangential velocity which exhibit the maximum reverse flow intensity relative to the boundary-layer edge velocity. The edge velocity of the local profiles was determined using the criterion of vanishing spanwise vorticity \textit{i.e.} $\overline{\omega}_{z}\approx0$. Since there is a very small but finite amount of vorticity in the far-field due to the free-stream turbulence, the criterion for boundary layer height is set as the wall-normal distance where the vorticity decays to $1\%$ of its maximum value in the boundary layer. Figure~\ref{fig:bubble_profiles} shows the wall-normal profiles of tangential velocity and spanwise vorticity along with the evaluated height of the boundary layer.

\begin{figure}
	\centering
	\begin{subfigure}[t]{0.48\textwidth}
		\centering
		\includegraphics[width=1\textwidth]{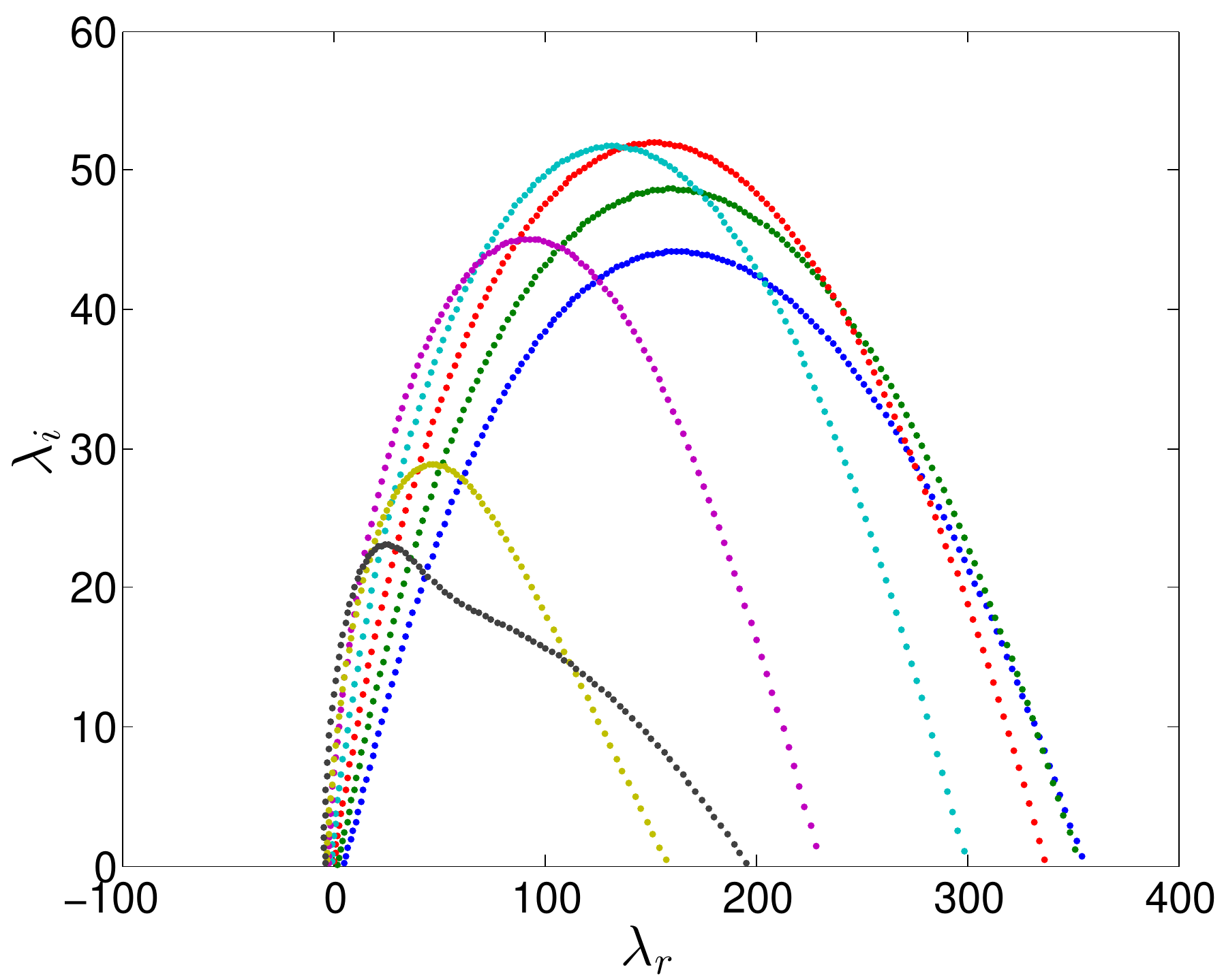}
	\end{subfigure}
	\begin{subfigure}[t]{0.48\textwidth}
		\centering
		\includegraphics[width=1\textwidth]{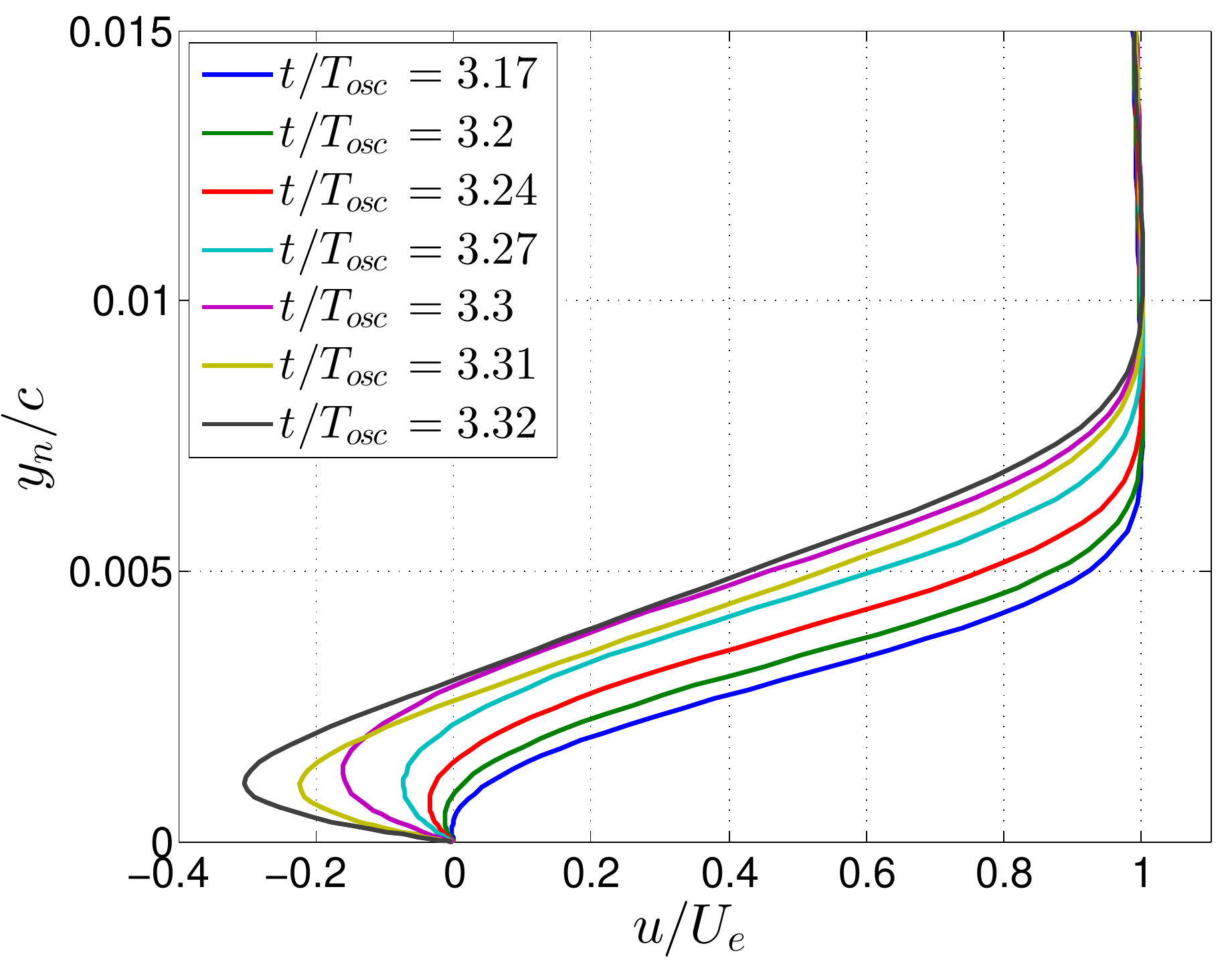}
	\end{subfigure}
	\caption{Unstable eigenvalues (left) obtained from a temporal stability analysis for different instantaneous velocity profiles (right).}
	\label{fig:bubble_lambda}
\end{figure}

\begin{figure}
	\centering
	\begin{subfigure}[t]{1\textwidth}
		\centering
		\includegraphics[width=0.9\textwidth]{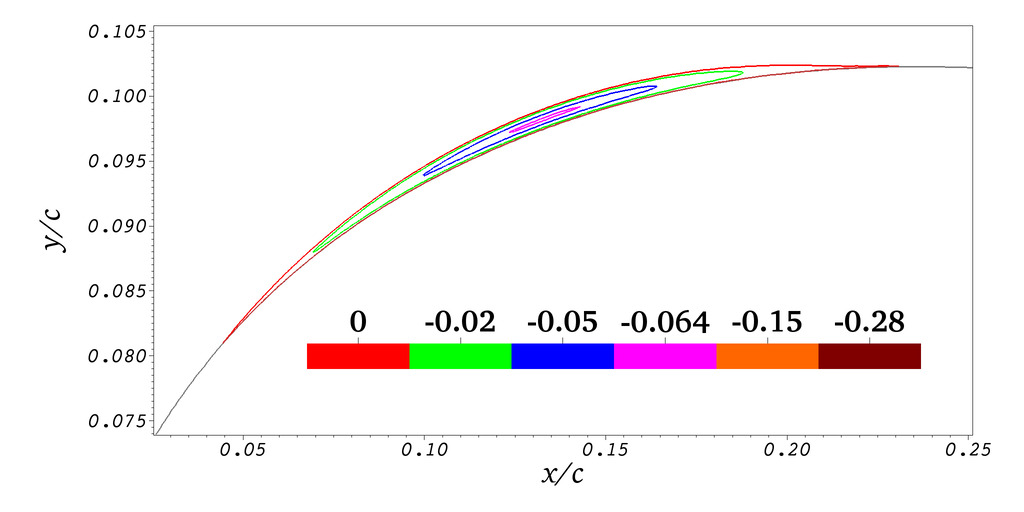}
	\end{subfigure}
	\begin{subfigure}[t]{1\textwidth}
		\centering
		\includegraphics[width=0.9\textwidth]{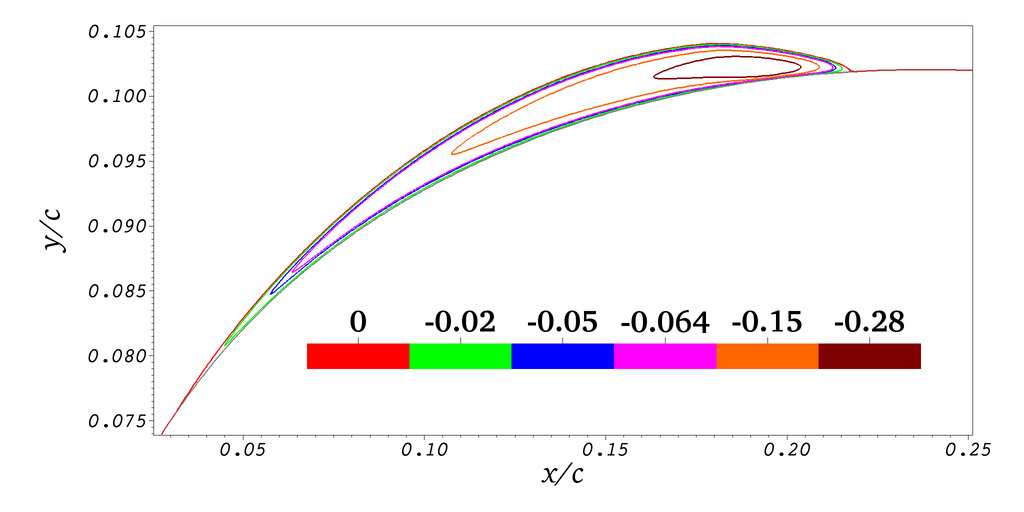}
	\end{subfigure}
	\caption{Contours of negative streamwise velocity in the leading-edge LSB at $t/T_{osc}=3.25$ (top) and $t/T_{osc}=3.32$ (bottom).}
	\label{fig:backflow_contours}
\end{figure}

Figure~\ref{fig:bubble_lambda} shows the unstable complex frequencies obtained from the temporal stability analysis (with varying $k_{x}$) for instantaneous profiles at several different time instants in the fourth pitch cycle. The reverse-flow intensity continues to increase with time until flow transition occurs at the LSB. The flow is unstable for all the analyzed velocity profiles as shown by the existence of complex frequencies with a positive imaginary part. However the highest amplification rate (frequency with maximum imaginary part) does not monotonically increase with reverse-flow intensity. At first the maximum amplification rate increases in time, but later it is seen to decrease. This changing characteristics can be associated with structural changes in the LSB, where at first the region of maximum reverse-flow is found near the center of the LSB, but as the LSB grows, this region of strong reverse flow moves closer to the downstream end of the LSB. Such qualitative changes in the shape of the LSB can be observed in figure~\ref{fig:backflow_contours}.

\begin{figure}
	\centering
	\includegraphics[width=0.8\textwidth]{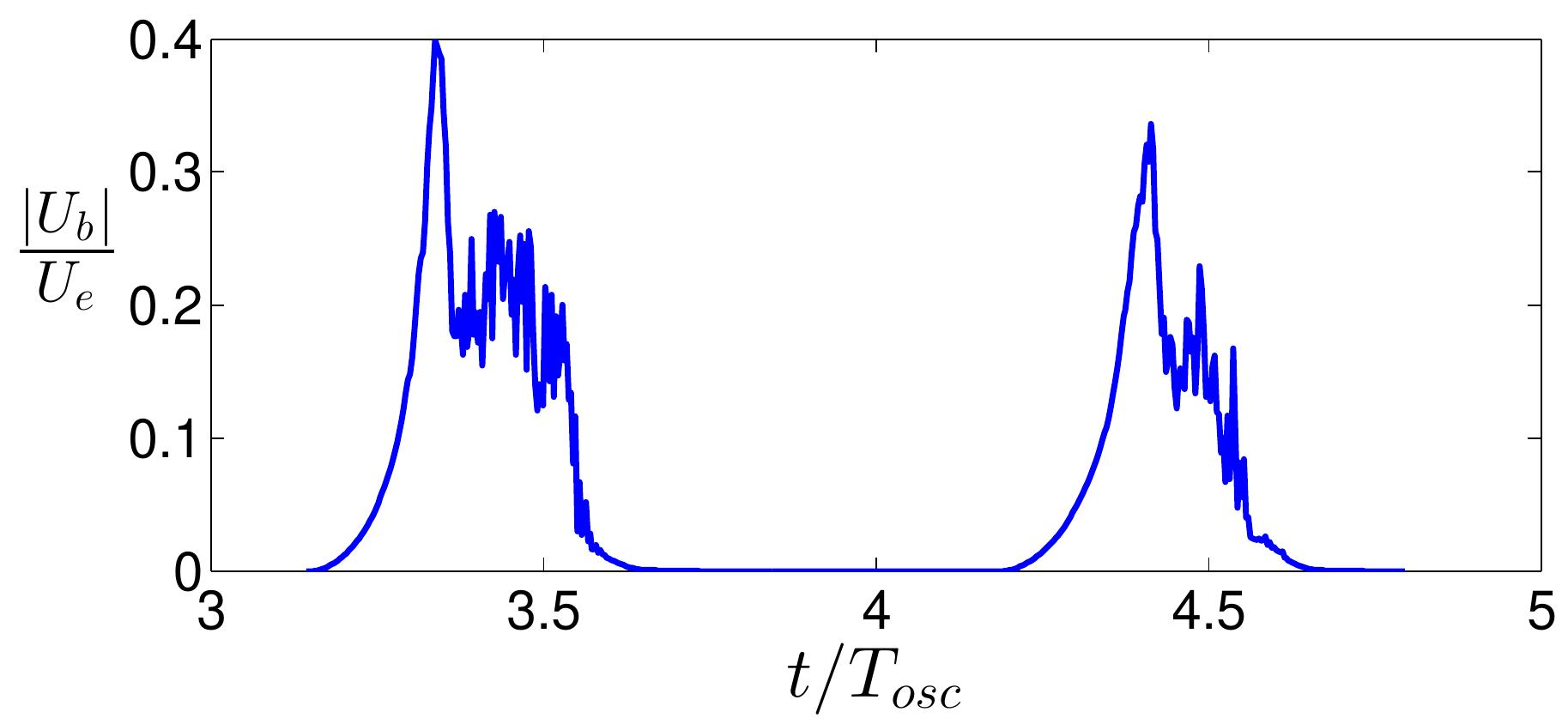}
	\caption{Ratio of maximum reverse flow ($U_{b}$) in the leading-edge LSB to the boundary layer edge velocity ($U_{e}$).} 
	\label{fig:backflow_ratio}
\end{figure}

While the local stability analysis shows that the boundary layer is unstable, an additional distinction needs to be made with regards to the nature of the instability which may be classified as either convective or absolute \citep{briggs64}. The instability characteristics are usually elucidated using the simple concept of group velocity of perturbations, $C_{g}$. Growing perturbations that travel with a positive group velocity are deemed convectively unstable since they move away from the source of disturbance. On the other hand, perturbations with zero group velocity are referred to as absolutely unstable, since they do not convect away from the origin of instability and create a self-sustaining process of perturbation growth. In the context of LSBs, high reverse-flow intensity has been associated with the presence of an absolute instability. \cite{alam00} with their local stability analysis of a two-parameter family of reverse flow profiles indicated that reverse flow intensities above $15\%$ may cause the flow to be locally absolutely unstable. With a similar analysis on a three parameter family of profiles \cite{hammond98} obtained onset of absolute instabilities at $20\%$ reverse flow velocities. In the same study the authors also performed global stability analysis on a synthetically created boundary layer with a symmetric separation bubble and found the flow to be globally unstable for $30\%$ reverse flow velocities. Figure~\ref{fig:backflow_ratio} shows the absolute value of the maximum reverse flow intensity in the leading-edge LSB found in the present study. The reverse flow velocities are found to be higher than $30\%$ in both the fourth and fifth pitch cycles, which is higher than the thresholds indicated by earlier investigators for absolute instability.
In such a case it is likely that the leading-edge LSB changes character to become absolutely unstable during the pitch cycle. The change of characteristics may explain the simultaneous existence of two different transition points in the boundary layer. Initially when the reverse-flow intensity in the LSB is small, the flow is convectively unstable and perturbations grow while traveling downstream. Thus transition occurs downstream of the LSB. As the reverse-flow intensity becomes larger the region of absolute instability may exist within the LSB which would cause perturbations to grow in time without being convected away. When these perturbations become large enough they would cause transition over the LSB. Thus momentarily, there would be two distinct transition points, one due to the growth of absolute instabilities in the LSB and one due to convectively amplifying disturbances which cause transition downstream of the LSB.

\begin{figure}
	\centering
	\begin{subfigure}[t]{0.48\textwidth}
		\centering
		\includegraphics[width=1\textwidth]{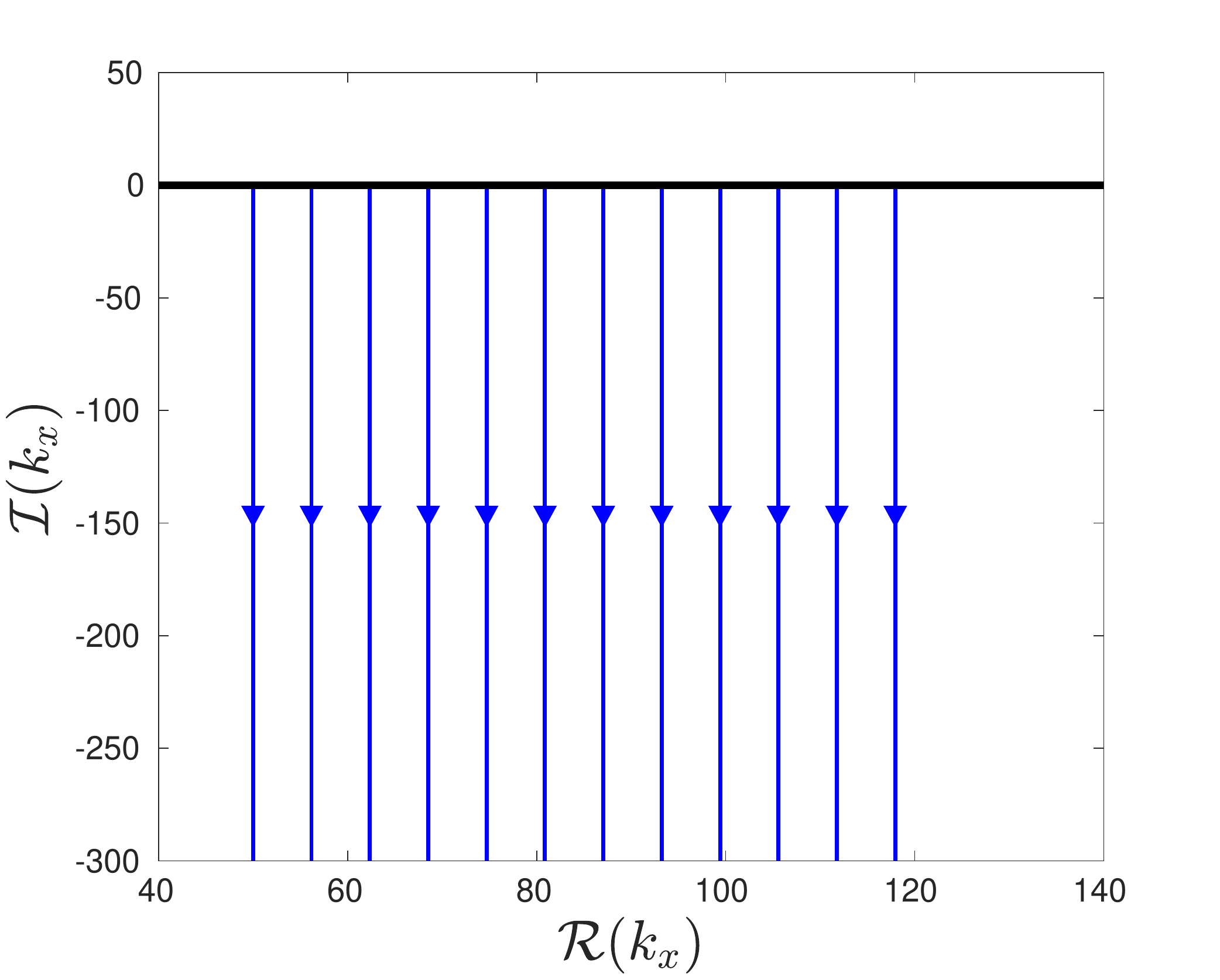}
	\end{subfigure}
	\begin{subfigure}[t]{0.48\textwidth}
		\centering
		\includegraphics[width=1\textwidth]{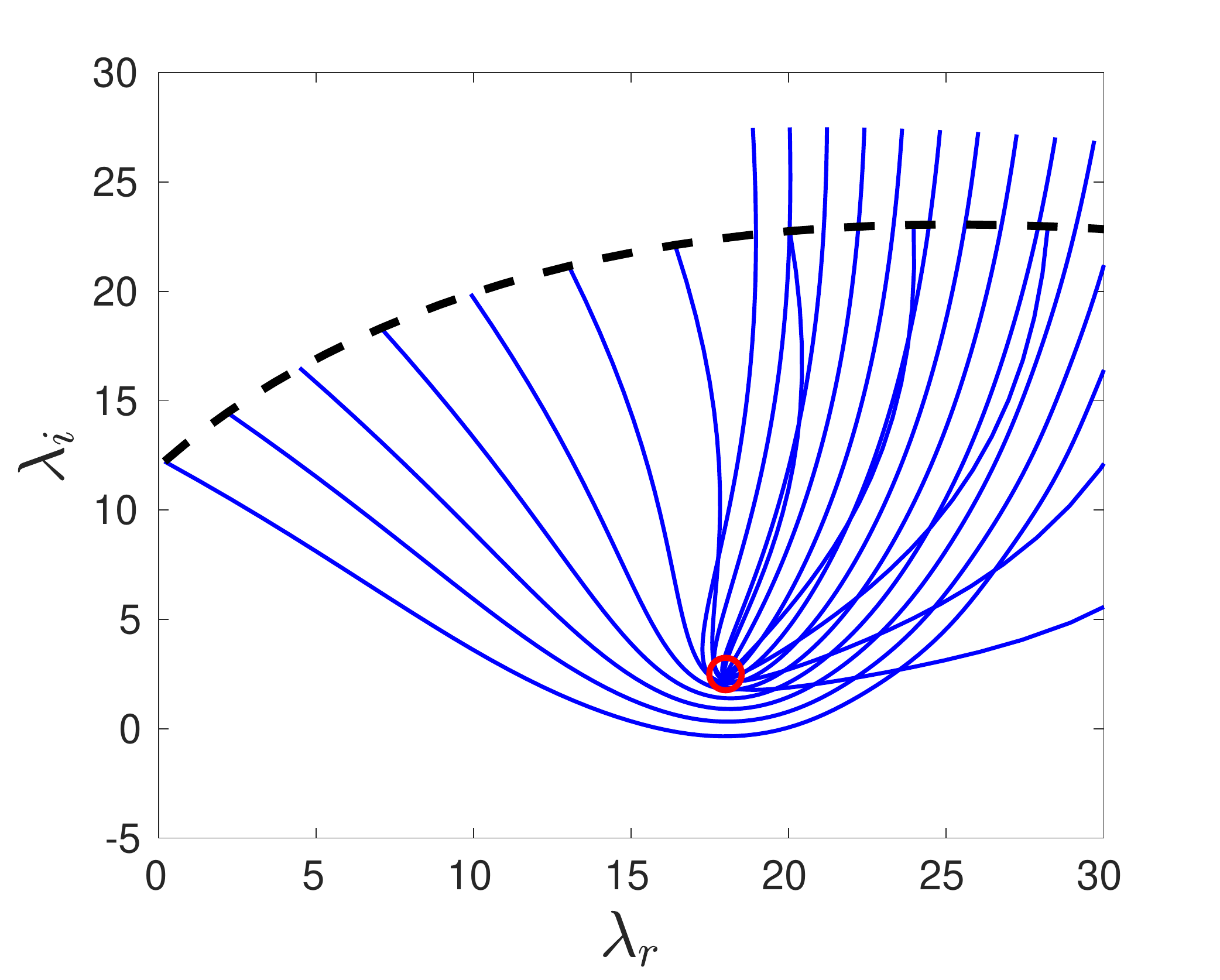}
	\end{subfigure}
	\caption{Contours on the complex wavenumber plane (left) and their corresponding mapping onto the complex frequency plane (right) at $t/T_{osc}=3.32$. Panel on the right shows the unstable cusp associated with absolute instability, located at $\lambda^{0}=18.02+2.46i$ (red circle). The dashed black line in the right panel corresponds to solutions of the Orr-Sommerfeld equation for real $k_{x}$ (black line on the left).}
	\label{fig:bubble_cusp_88}
\end{figure}

\begin{figure}
	\centering
	\begin{subfigure}[t]{0.48\textwidth}
		\centering
		\includegraphics[width=1\textwidth]{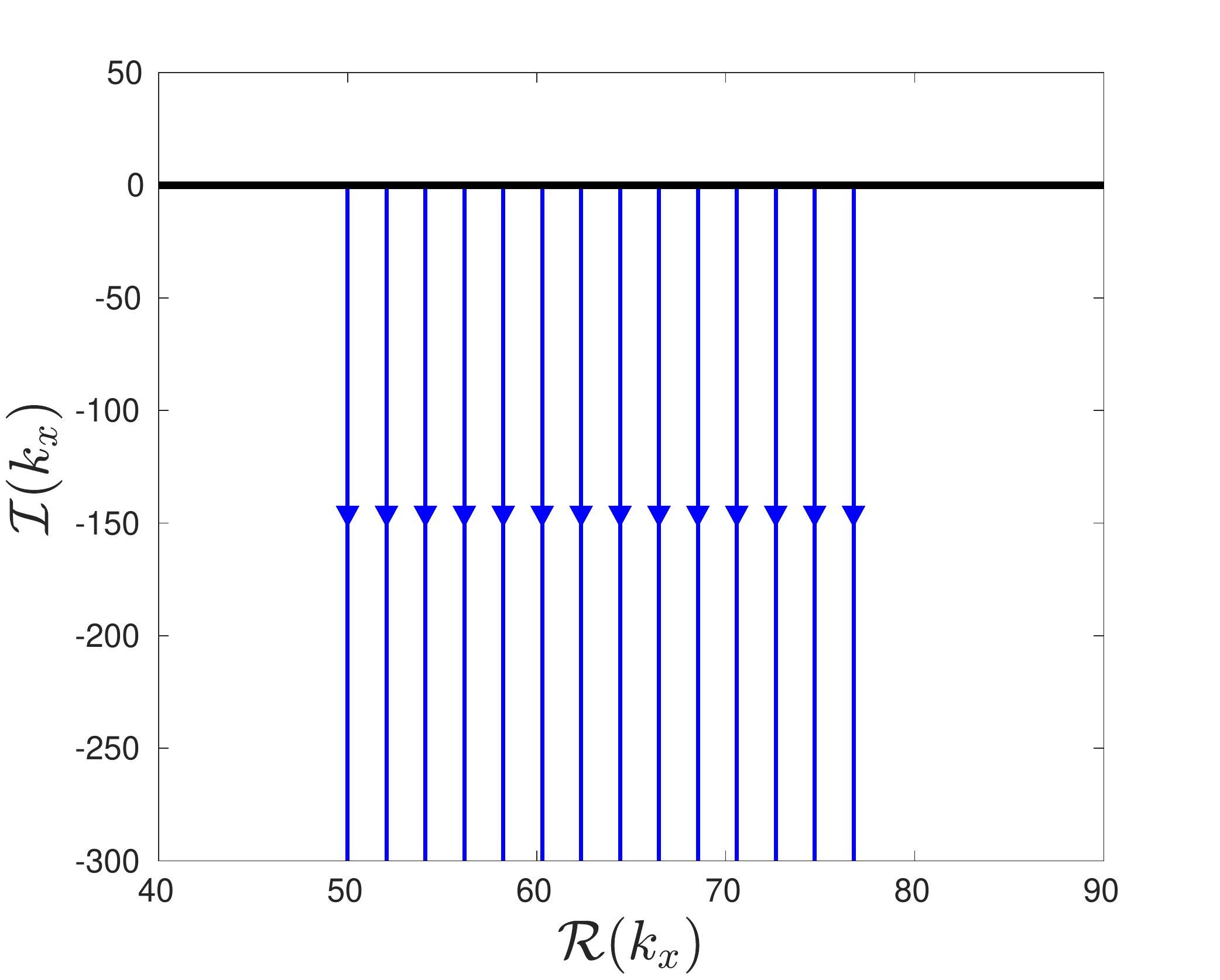}
	\end{subfigure}
	\begin{subfigure}[t]{0.48\textwidth}
		\centering
		\includegraphics[width=1\textwidth]{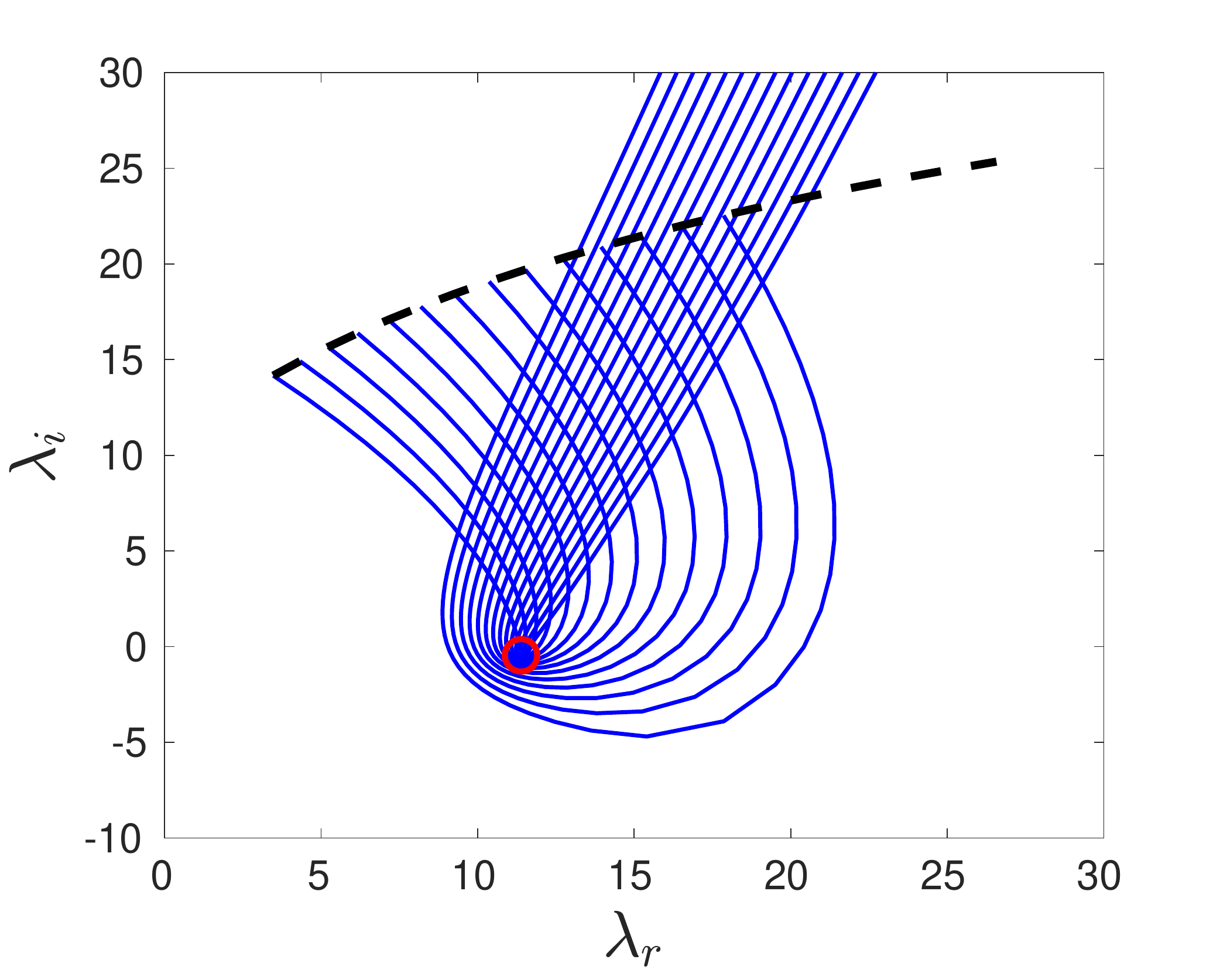}
	\end{subfigure}
	\caption{Contours on the complex wavenumber plane (left) and their corresponding mapping onto the complex frequency plane (right) at $t/T_{osc}=4.40$. Panel on the right shows the marginally stable cusp located at $\lambda^{0}=11.41-0.49i$ (red circle). }
	\label{fig:bubble_cusp_427}
\end{figure}

To explore such a possibility, from the local stability analysis, one needs to identify unstable modes with zero group velocity. \cite{briggs64} proposed a general method for the identification of absolute instabilities in a system which is commonly referred to as the Briggs method. The method involves solving the spatial stability problem for a range of complex $\lambda$, thus mapping contours on the complex frequency plane on to the complex wavenumber plane through the dispersion relation. Saddle points obtained in the complex wavenumber plane are the points where the dispersion relation has a double root. These points are known as the ``pinch-points'' in the complex wavenumber plane where the branches corresponding to the forward and backward propagating solutions of the dispersion relation meet via a double root. These pinch points correspond to perturbation modes that have zero group velocity. The mapping of the saddle-point on the frequency plane gives the absolute frequency $\lambda^{0}$. If this absolute frequency lies in the unstable half of the plane, then the system exhibits an absolute instability. \cite{kupfer87} proposed the inverse method called the cusp-map method, which maps contours in the complex wavenumber plane on to the complex frequency plane via the dispersion relation and identified the absolute instability by locating the cusp in the complex frequency plane. This allowed for solving the simpler linear eigenvalue problem for $\lambda$ rather than the non-linear eigenvalue problem for $k_{x}$ in the Briggs method \citep{briggs64}. Several contours need to be mapped from the wavenumber to the frequency plane for the location of the cusp. \cite{kupfer87} proposed mapping contours along lines with constant real part of $k_{x}$. Figure~\ref{fig:bubble_cusp_88} and \ref{fig:bubble_cusp_427} show the cusps obtained in the complex frequency plane for the profiles with the highest reverse-flow intensity of tangential velocity. In the fourth cycle at $t/T_{osc}=3.32$ an unstable cusp is obtained with a positive absolute amplification rate of $\lambda^{0}_{i}=2.46$. On the other hand the cusp found in the fifth cycle is just marginally stable with an absolute amplification rate of $\lambda^{0}_{i}=-0.5$ at $t/T_{osc}=4.4$. Since the velocity profiles considered here are averaged values instead of stationary base-flow profiles, it is possible that small temporal variation and/or non-linear modification of the flow may hide the absolute instability properties when an averaged flow-field is considered. Nonetheless the results provide a strong indication that the LSB changes character to become absolutely unstable during the pitch cycle.

\begin{figure}
	\centering
	\includegraphics[width=1\textwidth]{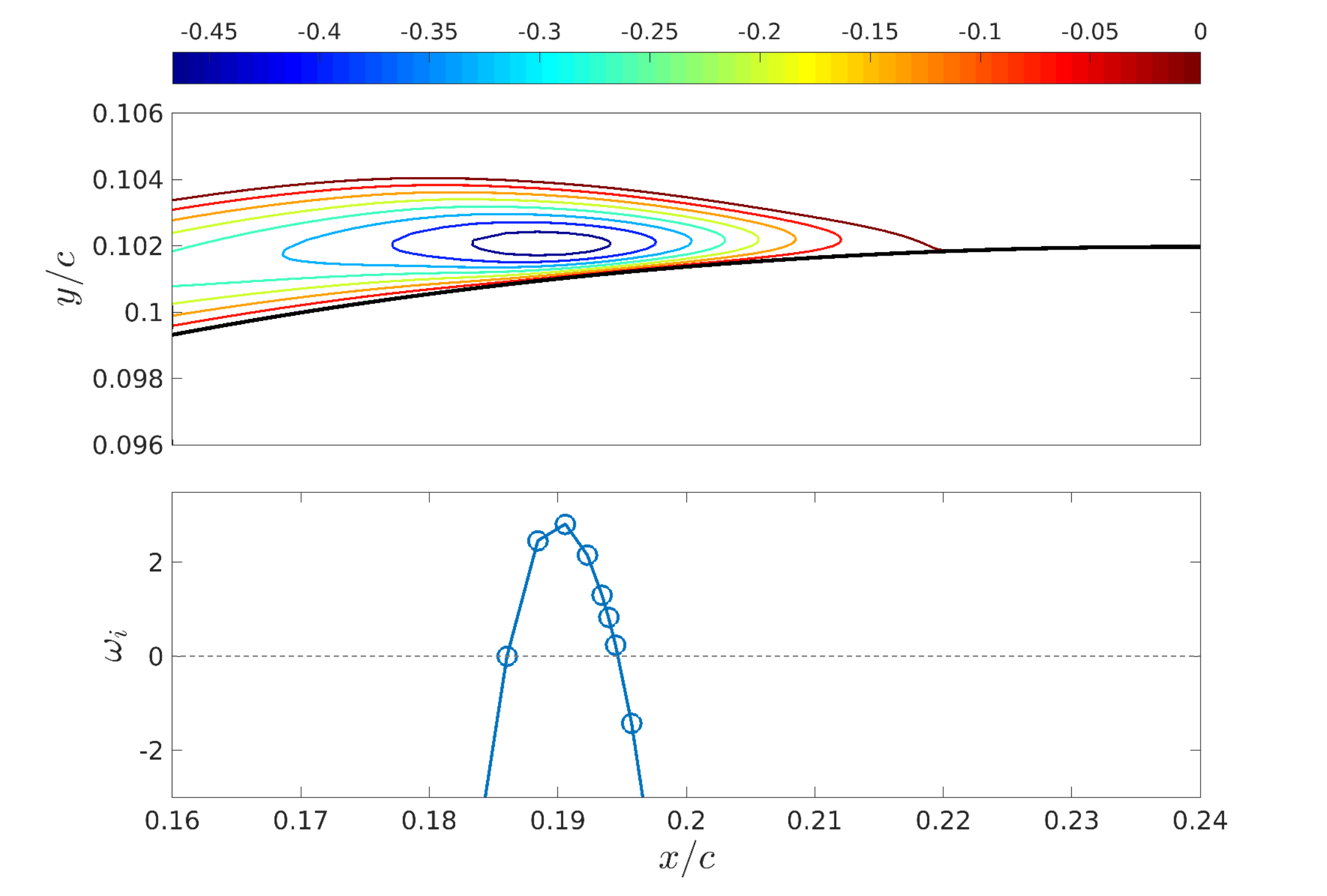}
	\caption{Top: Contours of negative stream-wise velocity near the leading-edge of the airfoil at $t/T_{osc}=3.32$. Bottom: Variation of absolute growth rate with the chord-wise location. (The $x$-axis is the same for both the panels.)}
	\label{fig:w0_x_88}
\end{figure}

According to \cite{chomaz91} and \cite{huerre90}, the existence of a finite region of absolute instability is a necessary criterion for flows to exhibit a global hydrodynamic instability. Local stability analysis is therefore performed at multiple chord-wise locations at $t/T_{osc}=3.32$. Figure~\ref{fig:w0_x_88} shows the spatial variation of the absolute growth rate along with the contours of negative streamwise velocity in the LSB. Clearly a small finite region of absolute instability exists close to the rear end of the LSB, where the highest intensities of reverse flow exist.

Caution needs to be taken in the interpretation of the stability analysis. The local stability analysis using the Orr-Sommerfeld equations assumes stationary homogeneous base state. Neither of which are strictly fulfilled when using the instantaneous, spanwise averaged velocity profiles. One can compare the properties of the LSB and the instability time-scales to judge the validity of the such an analysis. The ratio of the time-period of the absolute instability ($4^{th}$ cycle) to the time period of oscillation is $0.05$, suggesting that the boundary layer would appear nearly stationary to the amplifying disturbances. For the spatial inhomogeneity, the ratio of the maximum boundary layer height to the length of the separation bubble can be used as an indicator. This ratio is equal to $\delta^{max}/L_{x}=0.03$ which suggests a weak spatial inhomogeneity. Here $\delta^{max}$ is the maximum value of the boundary layer height over the LSB and $L_{x}$ is the spatial extent of the LSB. Both the above quantities indicate the quasi-steady, homogeneous flow assumptions may be used to obtain qualitative features of the flow case. The analysis then suggests that the convective instability properties of the boundary layer initially become stronger as the LSB grows. The LSB also changes in shape, with regions of high reverse-flow moving to the end of the bubble. At high reverse-flow intensities the LSB changes character and exhibits a region of absolute instability. This can potentially explain the emergence of two distinct transition locations. The upstream transition would be caused by the temporally growing instabilities which amplify within the region of absolute instability without being convected downstream. On the other hand spatially growing waves associated with convective instabilities would trigger transition further downstream of the LSB. The emergence of the second transition point associated with absolute instabilities would cause abrupt changes in the boundary-layer characteristics.

As mentioned earlier, due to conditions of weak spatial inhomogeneity and slow time-variation, the analysis of local stability analysis remains indicative rather than conclusive. Global instabilities may be triggered by mechanisms different from those arising due to an absolute instability \citep{huerre90}. Studies have claimed alternate mechanisms of global instability in laminar separation bubbles. \cite{cherubini10,theofilis00,rodriquez10} propose topological changes in the separated flow region as the cause of global instability. \cite{jones08,marxen13} provide some evidence that secondary instabilities of the vortices shed from the LSB can lead to a  self-sustaining transition. \cite{jones08} claim that a combination of hyperbolic instability in the braid region of vortices and elliptic instability of the vortex core causes the flow to behave like an oscillator. \cite{marxen13} also find evidence of elliptic and hyperbolic instabilities in their numerical study. In the current work we do not make a rigorous attempt to rule out all other mechanism of global instability. However some comments can be made on the possible existence of these other mechanisms in the current case. Topological changes, as proposed by \cite{theofilis00,rodriquez10} or the one by \cite{cherubini10} cause the LSB to get divided into two regions. Velocity contours inside the LSB for our case (figure~\ref{fig:backflow_contours}) show no such division of the LSB. Therefore such topological changes may not be the cause of transition in the present case. The flow cases studied by \cite{jones08,marxen13} rely on the secondary instability for sustained turbulence. However in both the cases the reverse flow intensities are small ($\sim12\%$) which ruled out the possibility of absolute instabilities in those cases. In contrast in the present case the LSB exhibits reverse flow intensities of $30-40\%$ which makes it substantially more likely for an absolute instability mechanism to be active.

\section{Conclusion}

A relaxation-term filtering procedure is used for wall-resolved LES of flow over a pitching airfoil. Validation of the LES procedure is done in a channel flow at $Re_{\tau}=395$ and for a wing section at $Re_{c}=400,000$ and the results show a good agreement with available DNS data sets.

Flow over an airfoil is simulated using the LES procedure at a chord based Reynolds number of $Re_{c}=100,000$ within an angle of attack range where the aerodynamic forces on the airfoil exhibit sensitive dependence on the angle of attack. This sensitive dependence is captured in the steady simulations at different angles of attack with large changes in transition location within a small variation of $\alpha$.

Pitch oscillations of the airfoil within this $\alpha$ range of sensitive dependence displays a rich variety of unsteady flow phenomena. The flow goes through alternating periods of fully turbulent and laminar flow over the suction-side of the airfoil with different governing mechanisms for transition through the oscillating phases. When the flow is mostly laminar over the airfoil surface it separates easily under the effect of adverse pressure gradient, forming an LSB near the trailing-edge. Flow transition occurs over this separated shear layer. As the angle of attack increases, a leading-edge LSB is formed which first excites spatially growing waves (convective instability) causing transition downstream of the LSB. Initially the amplification rate of these spatially growing waves increases as the size of the LSB grows causing transition to move upstream. Eventually a region of absolute instability develops within the LSB and flow transition occurs abruptly on the separated shear layer. When transition is first triggered by this absolute instability mechanism the flow exhibits two distinct transition locations and abrupt changes in the boundary layer follow.

In the pitch-down cycle, the reverse phenomenon occurs where the leading-edge LSB shrinks in size and the region of absolute instability ceases to exit. The transition is then again governed by spatially amplifying waves. The spatial amplification rate now reduces as the LSB shrinks and transition moves downstream. The flow thus goes through states of convective and absolute instability, leading to constantly changing transition location. The upstream and downstream velocities of the transition point movement however are vastly different, with an average upstream velocity being around $V^{tr}_{u}\approx-0.60$ and a much slower downstream velocity of $V^{tr}_{d}\approx0.17$. This asymmetry is yet to be investigated, but may be an important parameter in unsteady turbulence modeling.

\section*{Acknowledgement} 

The computations were performed on resources provided by the Swedish National Infrastructure for Computing (SNIC) at the PDC Center for High Performance Computing at the Royal Institute of Technology (KTH). Simulation have also been performed at the Barcelona Supercomputing Center, Barcelona, with computer time provided by the $12^{th}$ PRACE Project Access Call (number 2015133182). The work was partially funded by European Research Council under grant agreement 694452-TRANSEP-ERC-2015-AdG. The work was also partially funded by Vinnova through the NFFP project UMTAPS, with grant number 2014-00933. We would like to thank Dr. David Eller and Dr. Mikaela Lokatt for providing us with the NLF design and the numerous discussions on different aerodynamic aspects of the project.

\appendix
\section{Empirical transition location}
\label{app:A}
\begin{figure}
	\centering
	\includegraphics[width=0.49\textwidth]{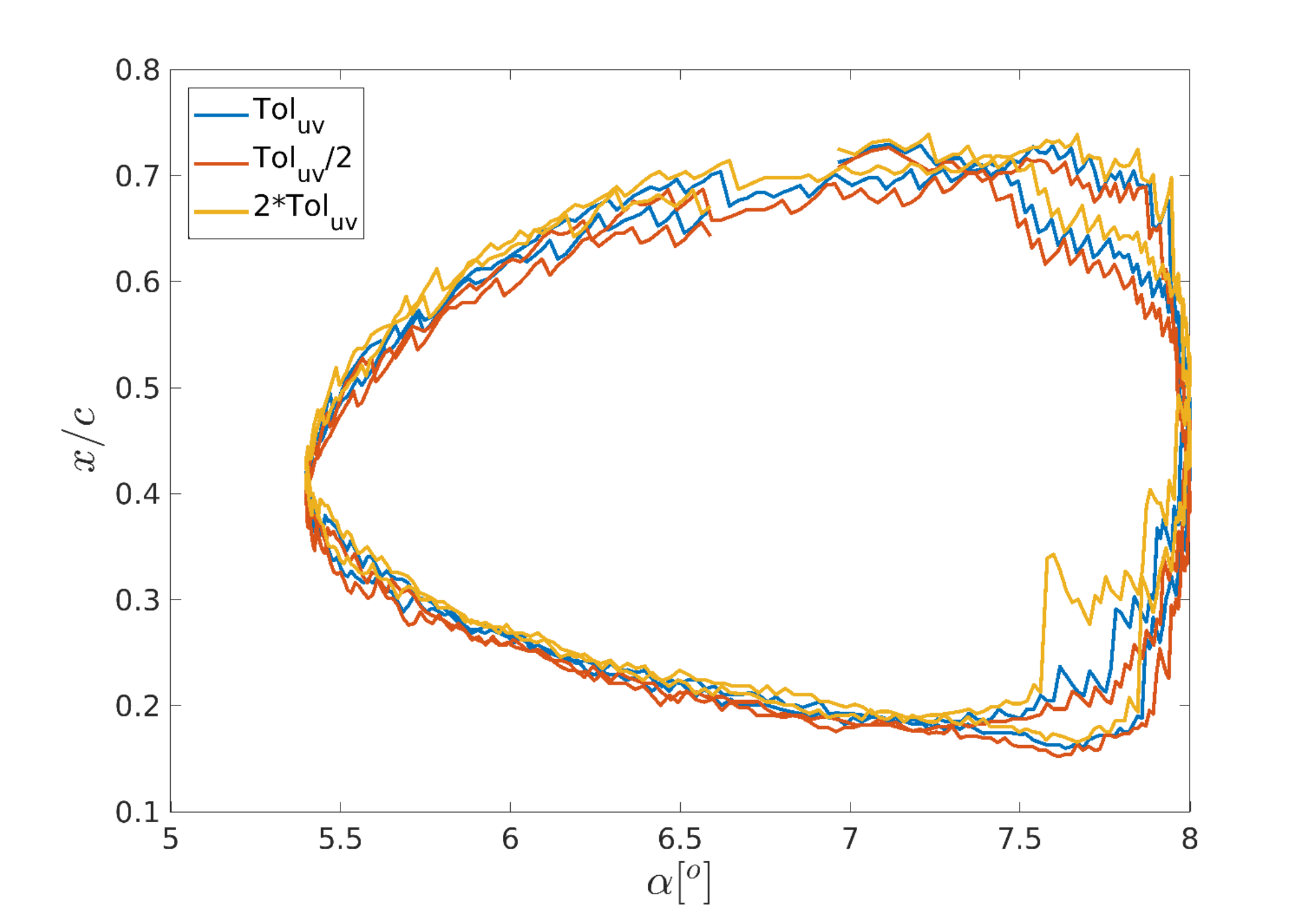}
	\includegraphics[width=0.49\textwidth]{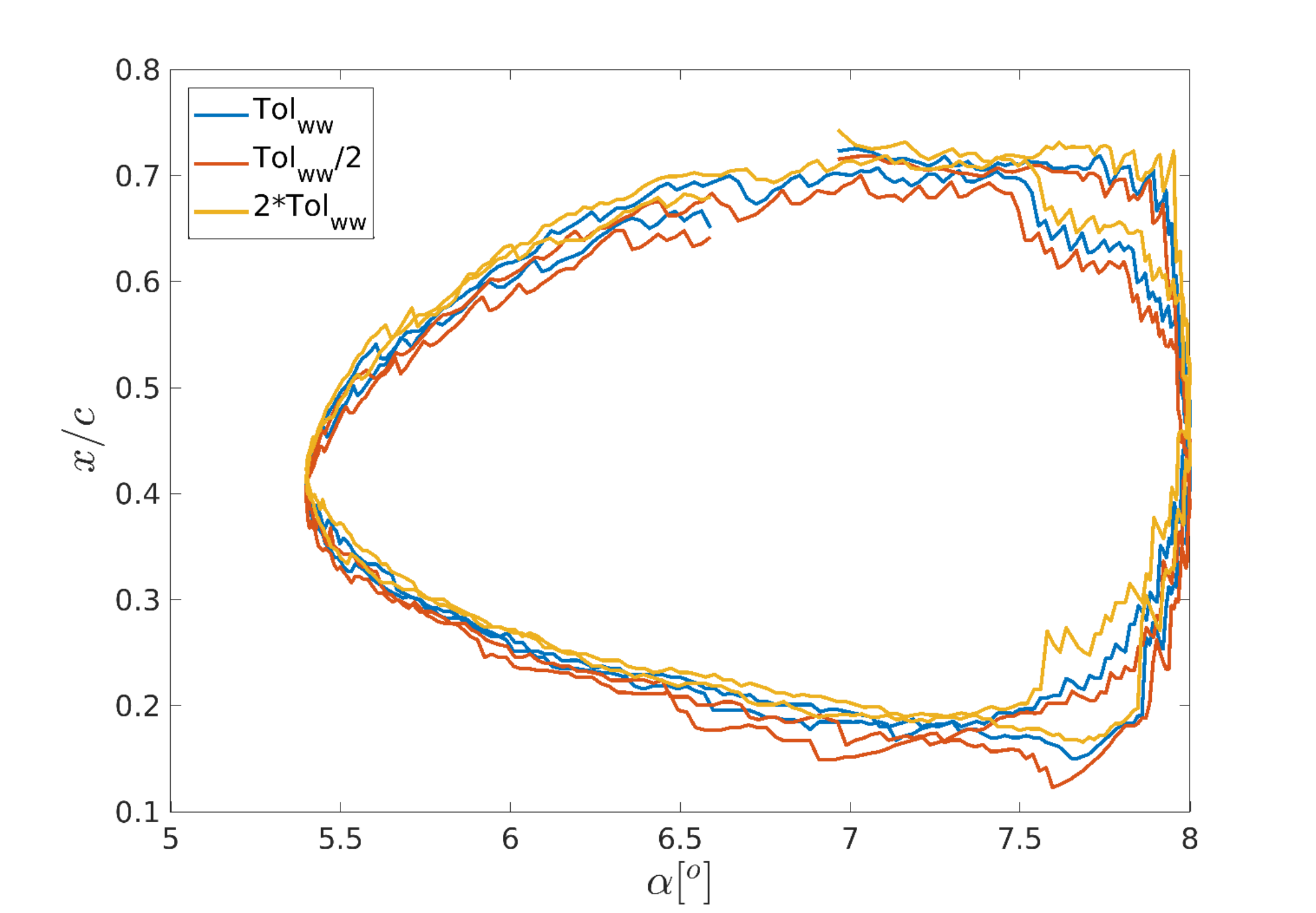}	
	\caption{Convergence of empirically determined transition locations. Left: Transition determined using $\overline{|u'v'|}$. Right: Transition determined using $|\overline{w'w'}|$}
	\label{fig:tr_conv}	
\end{figure}

In order to quantify the transition location, we calculate statistical quantities which are averaged in the homogeneous spanwise direction, and also averaged for a short duration ($\Delta t$) in time. This averaged value is considered as the ``instantaneous'' value for that quantity. Thus this instantaneous value of any statistical quantity $\overline{q}(x,y,t)$ can be evaluated as in equation~\ref{eqn:inst_q}:
\begin{align}
\overline{q}(x,y,t) = \left(\frac{1}{z_{max}-z_{min}}\right)\left(\frac{1}{\Delta t}\right)\int_{t'=t}^{t'=t+\Delta t}\int_{z=z_{min}}^{z=z_{max}}q(x,y,z,t')\ dz\ dt'.
\label{eqn:inst_q}
\end{align}
Here $(z_{max}-z_{min})$ is the spanwise extent of the computational domain. In order for such a quantity to be representative of the instantaneous state of the flow, the time duration of the averaging must be small. For the current case we use $\Delta t=4\times10^{-3}$, which amounts to $0.06\%$ of the oscillation time period during which the flow can be assumed to remain in approximately the same state. Using this procedure we evaluate the fluctuating Reynolds stress, $\overline{u'v'}(x,y,t)$, in order to determine the instantaneous transition location. The first streamwise location (on the suction-side) where the fluctuating Reynolds stress becomes larger than a certain threshold is denoted as the transition point. 
In order to prescribe a suitable threshold, the maximum value of $|\overline{u'v'}|$ across the entire boundary layer is evaluated for all times. This maximum value does not exhibit large variations, remaining within the same order of magnitude for all times with its mean value being $|\overline{u'v'}|_{max}\approx0.05$. The threshold for determining transition is set to $5\%$ of this value. The transition point is thus the first point where $|\overline{u'v'}|>Tol_{uv}=0.0025$. Since we use an ad-hoc criterion for transition location, this is cross-checked by evaluating the variance of the spanwise velocity fluctuations $\overline{w'w'}$, and following an identical procedure as described above. In this case the transition criterion is prescribed as $|\overline{w'w'}|>Tol_{ww}=0.005$, since the peak spanwise fluctuation intensities are found to be nearly twice the peak Reynolds stress $|\overline{u'v'}|$. Physically, growing spanwise velocity fluctuations indicate the onset of three-dimensionality in the boundary layer, which are associated with secondary instabilities. While the thresholds specified may be considered ad-hoc, the qualitative picture of transition movement is not very sensitive to small changes in the threshold. Changing the thresholds by a factor of 2 still produces the same qualitative trends as seen in figure~\ref{fig:tr_conv}. Moreover, the time variation of transition point determined by two different physical quantities agree very well with each other. Thus we consider the quantities as a good representations of instantaneous flow characteristics.

\FloatBarrier
\bibliographystyle{apa}

\begin{thebibliography}{}

\bibitem[\protect\astroncite{Alam and Sandham}{2000}]{alam00}
Alam, M. and Sandham, N.~D. (2000).
\newblock Direct numerical simulation of 'short' laminar separation bubbles
  with turbulent reattachment.
\newblock {\em Journal of Fluid Mechanics}, 410:1--28.

\bibitem[\protect\astroncite{Alferez et~al.}{2013}]{alferez13}
Alferez, N., Mary, I., and Lamballais, E. (2013).
\newblock Study of stall development around an airfoil by means of high
  fidelity large eddy simulation.
\newblock {\em Flow, Turbulence and Combustion}, 91(3):623--641.

\bibitem[\protect\astroncite{Barnes and Visbal}{2016}]{barnes16}
Barnes, C.~J. and Visbal, M. (2016).
\newblock High-fidelity les simulations of self-sustained pitching oscillations
  on a naca0012 airfoil at transitional reynolds numbers.
\newblock In {\em 54th AIAA Aerospace Sciences Meeting, AIAA SciTech Forum}.

\bibitem[\protect\astroncite{Barnes and Visbal}{2018}]{barnes18}
Barnes, C.~J. and Visbal, M.~R. (2018).
\newblock On the role of flow transition in laminar separation flutter.
\newblock {\em Journal of Fluids and Structures}, 77:213 -- 230.

\bibitem[\protect\astroncite{Boutilier and Yarusevych}{2012}]{boutilier12}
Boutilier, M. S.~H. and Yarusevych, S. (2012).
\newblock Separated shear layer transition over an airfoil at a low reynolds
  number.
\newblock {\em Physics of Fluids}, 24(8):084105.

\bibitem[\protect\astroncite{Brandt et~al.}{2004}]{brandt04}
Brandt, L., Schlatter, P., and Henningson, D.~S. (2004).
\newblock Transition in boundary layers subject to free-stream turbulence.
\newblock {\em Journal of Fluid Mechanics}, 517.

\bibitem[\protect\astroncite{Briggs}{1964}]{briggs64}
Briggs, R.~J. (1964).
\newblock {\em {Electron-Stream Interaction with Plasmas}}.
\newblock The MIT Press, Cambridge, Massachusetts.

\bibitem[\protect\astroncite{Carr et~al.}{1977}]{carr1977}
Carr, L.~W., McAlister, K.~W., and McCroskey, W.~J. (1977).
\newblock Analysis of the development of dynamic stall based on oscillating
  airfoil experiments.
\newblock Technical report, NASA Ames Research Center; Moffett Field, CA,
  United States.

\bibitem[\protect\astroncite{Cherubini et~al.}{2010}]{cherubini10}
Cherubini, S., Robinet, J.-C., and Palma, P.~D. (2010).
\newblock The effects of non-normality and nonlinearity of the navier--stokes
  operator on the dynamics of a large laminar separation bubble.
\newblock {\em Physics of Fluids}, 22(1):014102.

\bibitem[\protect\astroncite{Chin et~al.}{2015}]{chin15}
Chin, C., Ng, H., Blackburn, H., Monty, J., and Ooi, A. (2015).
\newblock Turbulent pipe flow at $\text{R}e_{\tau}= 1000$: A comparison of
  wall-resolved large-eddy simulation, direct numerical simulation and hot-wire
  experiment.
\newblock {\em Computers and Fluids}, 122:26 -- 33.

\bibitem[\protect\astroncite{Chomaz et~al.}{1991}]{chomaz91}
Chomaz, J.-M., Huerre, P., and Redekopp, L.~G. (1991).
\newblock A frequency selection criterion in spatially developing flows.
\newblock {\em Studies in Applied Mathematics}, 84(2):119--144.

\bibitem[\protect\astroncite{Choudhry et~al.}{2014}]{choudhry14}
Choudhry, A., Leknys, R., Arjomandi, M., and Kelso, R. (2014).
\newblock An insight into the dynamic stall lift characteristics.
\newblock {\em Experimental Thermal and Fluid Science}, 58:188 -- 208.

\bibitem[\protect\astroncite{Coorke and Thomas}{2015}]{coorke15}
Coorke, T.~C. and Thomas, F.~O. (2015).
\newblock Dynamic stall in pitching airfoils: Aerodynamic damping and
  compressibility effects.
\newblock {\em Annual Review of Fluid Mechanics}, 47(1):479--505.

\bibitem[\protect\astroncite{Dong et~al.}{2014}]{dong2014}
Dong, S., Karniadakis, G.~E., and Chryssostomidis, C. (2014).
\newblock A robust and accurate outflow boundary condition for incompressible
  flow simulations on severely-truncated unbounded domains.
\newblock {\em Journal of Computational Physics}, 261:83--105.

\bibitem[\protect\astroncite{Drela}{1989}]{drela89}
Drela, M. (1989).
\newblock {\em XFOIL: An Analysis and Design System for Low Reynolds Number
  Airfoils}, pages 1--12.
\newblock Springer Berlin Heidelberg, Berlin, Heidelberg.

\bibitem[\protect\astroncite{Dunne and McKeon}{2015}]{dunne2015}
Dunne, R. and McKeon, B.~J. (2015).
\newblock Dynamic stall on a pitching and surging airfoil.
\newblock {\em Experiments in Fluids}, 56(8):157.

\bibitem[\protect\astroncite{Eitel-Amor et~al.}{2014}]{eitel14}
Eitel-Amor, G., \"{O}rl\"{u} R., and Schlatter, P. (2014).
\newblock Simulation and validation of a spatially evolving turbulent boundary
  layer up to $\text{R}e_{\theta}=8300$.
\newblock {\em International Journal of Heat and Fluid Flow}, 47:57--69.

\bibitem[\protect\astroncite{Ericsson and Reding}{1988a}]{ericsson_stall88a}
Ericsson, L. and Reding, J. (1988a).
\newblock Fluid mechanics of dynamic stall part i. unsteady flow concepts.
\newblock {\em Journal of Fluids and Structures}, 2(1):1 -- 33.

\bibitem[\protect\astroncite{Ericsson and Reding}{1988b}]{ericsson_stall88b}
Ericsson, L. and Reding, J. (1988b).
\newblock Fluid mechanics of dynamic stall part ii. prediction of full scale
  characteristics.
\newblock {\em Journal of Fluids and Structures}, 2(2):113 -- 143.

\bibitem[\protect\astroncite{Fischer et~al.}{2008}]{nek5000}
Fischer, P.~F., Lottes, J.~W., and Kerkemeier, S.~G. (2008).
\newblock Nek5000 web page.
\newblock \url{http://nek5000.mcs.anl.gov}.

\bibitem[\protect\astroncite{H\"aggmark et~al.}{2001}]{haggmark01b}
H\"aggmark, C.~P., Hildings, C., and Henningson, D.~S. (2001).
\newblock A numerical and experimental study of a transitional separation
  bubble.
\newblock {\em Aerospace Science and Technology}, 5(5):317--328.

\bibitem[\protect\astroncite{Halfman}{1952}]{halfman52}
Halfman, R.~L. (1952).
\newblock Experimental aerodynamic derivatives of a sinusoidally oscillating
  airfoil in two-dimensional flow.
\newblock Technical report, National Advisory Committee for Aeronautics;
  Washington, DC, United States.

\bibitem[\protect\astroncite{Hammond and Redekopp}{1998}]{hammond98}
Hammond, D. and Redekopp, L. (1998).
\newblock Local and global instability properties of separation bubbles.
\newblock {\em European Journal of Mechanics - B/Fluids}, 17(2):145 -- 164.

\bibitem[\protect\astroncite{Hebler et~al.}{2013}]{hebler13}
Hebler, A., Schojda, L., and Mai, H. (2013).
\newblock Experimental investigation of the aeroelastic behavior of a laminar
  airfoil in transonic flow.
\newblock In {\em Proceedings IFASD}.

\bibitem[\protect\astroncite{Ho and Patera}{1990}]{ho90}
Ho, L.-W. and Patera, A.~T. (1990).
\newblock {A Legendre spectral element method for simulation of unsteady
  incompressible viscous free-surface flows}.
\newblock {\em Computer Methods in Applied Mechanics and Engineering},
  80(1):355 -- 366.

\bibitem[\protect\astroncite{Ho and Patera}{1991}]{ho91}
Ho, L.-W. and Patera, A.~T. (1991).
\newblock Variational formulation of three-dimensional viscous free-surface
  flows: Natural imposition of surface tension boundary conditions.
\newblock {\em International Journal for Numerical Methods in Fluids},
  13(6):691--698.

\bibitem[\protect\astroncite{Hosseini et~al.}{2016}]{hosseini16}
Hosseini, S.~M., Vinuesa, R., Schlatter, P., Hanifi, A., and Henningson, D.~S.
  (2016).
\newblock Direct numerical simulation of the flow around a wing section at
  moderate {R}eynolds number.
\newblock {\em International Journal of Heat and Fluid Flow}, 61:117 -- 128.

\bibitem[\protect\astroncite{Huerre and Monkewitz}{1990}]{huerre90}
Huerre, P. and Monkewitz, P.~A. (1990).
\newblock Local and global instabilities in spatially developing flows.
\newblock {\em Annual Review of Fluid Mechanics}, 22(1):473--537.

\bibitem[\protect\astroncite{Jeong and Hussain}{1995}]{jeong95}
Jeong, J. and Hussain, F. (1995).
\newblock On the identification of a vortex.
\newblock {\em Journal of Fluid Mechanics}, 285.

\bibitem[\protect\astroncite{Jones et~al.}{2008}]{jones08}
Jones, L.~E., Sandberg, R.~D., and Sandham, N.~D. (2008).
\newblock Direct numerical simulations of forced and unforced separation
  bubbles on an airfoil at incidence.
\newblock {\em Journal of Fluid Mechanics}, 602:175--207.

\bibitem[\protect\astroncite{Jones et~al.}{2010}]{jones10}
Jones, L.~E., Sandberg, R.~D., and Sandham, N.~D. (2010).
\newblock Stability and receptivity characteristics of a laminar separation
  bubble on an aerofoil.
\newblock {\em Journal of Fluid Mechanics}, 648:257--296.

\bibitem[\protect\astroncite{Kleusberg}{2017}]{kleusberglicenciate}
Kleusberg, E. (2017).
\newblock {\em Wind turbine simulations using spectral elements}.
\newblock Licentiate thesis, Royal Institute of Technology (KTH), Stockholm,
  Sweden.

\bibitem[\protect\astroncite{Kupfer et~al.}{1987}]{kupfer87}
Kupfer, K., Bers, A., and Ram, A.~K. (1987).
\newblock The cusp map in the complex-frequency plane for absolute
  instabilities.
\newblock {\em The Physics of Fluids}, 30(10):3075--3082.

\bibitem[\protect\astroncite{Lang et~al.}{2004}]{lang04}
Lang, M., Rist, U., and Wagner, S. (2004).
\newblock Investigations on controlled transition development in a laminar
  separation bubble by means of lda and piv.
\newblock {\em Experiments in Fluids}, 36(1):43--52.

\bibitem[\protect\astroncite{Langtry and Menter}{2009}]{langtry09}
Langtry, R.~B. and Menter, F.~R. (2009).
\newblock {Correlation-Based Transition Modeling for Unstructured Parallelized
  Computational Fluid Dynamics Codes}.
\newblock {\em AIAA Journal}, 47:2894--2906.

\bibitem[\protect\astroncite{Leishman}{2000}]{leishman00}
Leishman, J.~G. (2000).
\newblock {\em Principles of Helicopter Aerodynamics}.
\newblock Cambridge University Press.

\bibitem[\protect\astroncite{Lokatt}{2017}]{lokattthesis}
Lokatt, M. (2017).
\newblock {\em On Aerodynamic and Aeroelastic Modeling for Aircraft Design}.
\newblock Doctoral thesis, KTH Royal Institute of Technology.

\bibitem[\protect\astroncite{Lokatt and Eller}{2017}]{lokatt17}
Lokatt, M. and Eller, D. (2017).
\newblock Robust viscous-inviscid interaction scheme for application on
  unstructured meshes.
\newblock {\em Computers $\&$ Fluids}, 145:37 -- 51.

\bibitem[\protect\astroncite{Lombard et~al.}{2016}]{lombard15}
Lombard, J.-E.~W., Moxey, D., Sherwin, S.~J., Hoessler, J.~F.~A., Dhandapani,
  S., and Taylor, M.~J. (2016).
\newblock {Implicit Large-Eddy Simulation of a Wingtip Vortex}.
\newblock {\em AIAA Journal}, 54:506--518.

\bibitem[\protect\astroncite{{Maday} and {Patera}}{1989}]{maday89}
{Maday}, Y. and {Patera}, A.~T. (1989).
\newblock {Spectral element methods for the incompressible Navier-Stokes
  equations}.
\newblock In {\em State-of-the-art surveys on computational mechanics
  (A90-47176 21-64). New York, American Society of Mechanical Engineers, 1989,
  p. 71-143. Research supported by DARPA.}, pages 71--143.

\bibitem[\protect\astroncite{Mai and Hebler}{2011}]{mai11}
Mai, H. and Hebler, A. (2011).
\newblock Aeroelasticity of a laminar wing.
\newblock In {\em Proceedings IFASD}, Paris.

\bibitem[\protect\astroncite{Marxen et~al.}{2012}]{marxen12}
Marxen, O., Lang, M., and Rist, U. (2012).
\newblock Discrete linear local eigenmodes in a separating laminar boundary
  layer.
\newblock {\em Journal of Fluid Mechanics}, 711:1--26.

\bibitem[\protect\astroncite{Marxen et~al.}{2013}]{marxen13}
Marxen, O., Lang, M., and Rist, U. (2013).
\newblock Vortex formation and vortex breakup in a laminar separation bubble.
\newblock {\em Journal of Fluid Mechanics}, 728:58--90.

\bibitem[\protect\astroncite{Marxen et~al.}{2003}]{marxen03}
Marxen, O., Lang, M., Rist, U., and Wagner, S. (2003).
\newblock A combined experimental/numerical study of unsteady phenomena in a
  laminar separation bubble.
\newblock {\em Flow, Turbulence and Combustion}, 71(1):133--146.

\bibitem[\protect\astroncite{Marxen and Rist}{2010}]{marxen10}
Marxen, O. and Rist, U. (2010).
\newblock Mean flow deformation in a laminar separation bubble: separation and
  stability characteristics.
\newblock {\em Journal of Fluid Mechanics}, 660:37--54.

\bibitem[\protect\astroncite{McCroskey}{1973}]{mccroskey73}
McCroskey, W.~J. (1973).
\newblock Inviscid flowfield of an unsteady airfoil.
\newblock {\em AIAA Journal}, 11(8):1130 -- 1137.

\bibitem[\protect\astroncite{McCroskey}{1981}]{mccroskey81}
McCroskey, W.~J. (1981).
\newblock Phenomenon of dynamic stall.
\newblock Technical report, NASA Ames Research Center; Moffett Field, CA,
  United States.

\bibitem[\protect\astroncite{McCroskey}{1982}]{mccroskey82}
McCroskey, W.~J. (1982).
\newblock Unsteady airfoils.
\newblock {\em Annual Review of Fluid Mechanics}, 14(1):285--311.

\bibitem[\protect\astroncite{McCroskey et~al.}{1976}]{mccroskey76}
McCroskey, W.~J., Carr, L.~W., and McAlister, K.~W. (1976).
\newblock Dynamic stall experiments on oscillating airfoils.
\newblock {\em AIAA Journal}, 14(1):57 -- 63.

\bibitem[\protect\astroncite{McCroskey et~al.}{1982}]{mccroskey82experimental}
McCroskey, W.~J., McAlister, K.~W., Carr, L.~W., and Pucci, S.~L. (1982).
\newblock An experimental study of dynamic stall on advanced airfoil sections.
  volume 1: Summary of the experiment.
\newblock Technical report, NASA Ames Research Center, Moffett Field, CA,
  United States.

\bibitem[\protect\astroncite{Moser et~al.}{1999}]{moser99}
Moser, R.~D., Kim, J., and Mansour, N.~N. (1999).
\newblock Direct numerical simulation of turbulent channel flow up to
  $\text{R}e_{\tau}=590$.
\newblock {\em Physics of Fluids}, 11(4):943--945.

\bibitem[\protect\astroncite{Nati et~al.}{2015}]{nati15}
Nati, A., De~Kat, R., Scarano, F., and Van~Oudheusden, B.~W. (2015).
\newblock Dynamic pitching effect on a laminar separation bubble.
\newblock {\em Experiments in Fluids}, 56(9):172.

\bibitem[\protect\astroncite{Pascazio et~al.}{1996}]{pascazio96}
Pascazio, M., Autric, J., Favier, D., and Maresca, C. (1996).
\newblock Unsteady boundary-layer measurement on oscillating
  airfoils-transition and separation phenomena in pitching motion.
\newblock In {\em 34th Aerospace Sciences Meeting and Exhibit}, page~35.

\bibitem[\protect\astroncite{Poirel et~al.}{2008}]{poirel08}
Poirel, D., Harris, Y., and Benaissa, A. (2008).
\newblock Self-sustained aeroelastic oscillations of a naca0012 airfoil at
  low-to-moderate reynolds numbers.
\newblock {\em Journal of Fluids and Structures}, 24(5):700 -- 719.

\bibitem[\protect\astroncite{Poirel and Yuan}{2010}]{poirel10}
Poirel, D. and Yuan, W. (2010).
\newblock Aerodynamics of laminar separation flutter at a transitional reynolds
  number.
\newblock {\em Journal of Fluids and Structures}, 26(7):1174 -- 1194.

\bibitem[\protect\astroncite{Rainey}{1957}]{rainey57}
Rainey, A.~G. (1957).
\newblock Measurement of aerodynamic forces for various mean angles of attack
  on an airfoil oscillating in pitch and on two finite-span wings oscillating
  in bending with emphasis on damping in the stall.
\newblock Technical report, National Advisory Committee for Aeronautics.
  Langley Aeronautical Lab.; Langley Field, VA, United States.

\bibitem[\protect\astroncite{Rival and Tropea}{2010}]{rival2010}
Rival, D. and Tropea, C. (2010).
\newblock Characteristics of pitching and plunging airfoils under dynamic-stall
  conditions.
\newblock {\em Journal of Aircraft}, 47(1):80--86.

\bibitem[\protect\astroncite{Rodr\'{i}guez and Theofilis}{2010}]{rodriquez10}
Rodr\'{i}guez, D. and Theofilis, V. (2010).
\newblock Structural changes of laminar separation bubbles induced by global
  linear instability.
\newblock {\em Journal of Fluid Mechanics}, 655:280--305.

\bibitem[\protect\astroncite{Rosti et~al.}{2016}]{rosti16}
Rosti, M.~E., Omidyeganeh, M., and Pinelli, A. (2016).
\newblock Direct numerical simulation of the flow around an aerofoil in ramp-up
  motion.
\newblock {\em Physics of Fluids}, 28(2):025106.

\bibitem[\protect\astroncite{Schlatter}{2001}]{schlatterdiploma}
Schlatter, P. (2001).
\newblock {\em Direct numerical simulation of laminar-turbulent transition in
  boundary layer subject to free-stream turbulence}.
\newblock Diploma thesis, Royal Institute of Technology (KTH), Stockholm,
  Sweden.

\bibitem[\protect\astroncite{Schlatter et~al.}{2008}]{schlatter08}
Schlatter, P., Brandt, L., de~Lange, H.~C., and Henningson, D.~S. (2008).
\newblock On streak breakdown in bypass transition.
\newblock {\em Physics of Fluids}, 20(10):101505.

\bibitem[\protect\astroncite{Schlatter et~al.}{2004}]{schlatter04}
Schlatter, P., Stolz, S., and Kleiser, L. (2004).
\newblock \text{LES} of transitional flows using the approximate deconvolution
  model.
\newblock {\em International Journal of Heat and Fluid Flow}, 25(3):549 -- 558.
\newblock Turbulence and Shear Flow Phenomena (TSFP-3).

\bibitem[\protect\astroncite{Schlatter et~al.}{2006a}]{schlatter06b}
Schlatter, P., Stolz, S., and Kleiser, L. (2006a).
\newblock {\em Analysis of the SGS energy budget for deconvolution- and
  relaxation-based models in channel flow}, pages 135--142.
\newblock Springer Netherlands, Dordrecht.

\bibitem[\protect\astroncite{Schlatter et~al.}{2006b}]{schlatter06}
Schlatter, P., Stolz, S., and Kleiser, L. (2006b).
\newblock Large-eddy simulation of spatial transition in plane channel flow.
\newblock {\em Journal of Turbulence}, 7:N33.

\bibitem[\protect\astroncite{Schmid and Henningson}{2001}]{schmid01}
Schmid, P.~J. and Henningson, D.~S. (2001).
\newblock {\em {Stability and Transition in Shear Flows}}.
\newblock Springer.

\bibitem[\protect\astroncite{Theodorsen}{1935}]{theodorsen35}
Theodorsen, T. (1935).
\newblock General theory of aerodynamic instability and the mechanism of
  flutter.
\newblock Technical report, National Advisory Committee for Aeronautics;
  Langley Aeronautical Lab.; Langley Field, VA, United States.

\bibitem[\protect\astroncite{Theofilis et~al.}{2000}]{theofilis00}
Theofilis, V., Hein, S., and Dallmann, U. (2000).
\newblock On the origins of unsteadiness and three-dimensionality in a laminar
  separation bubble.
\newblock {\em Philosophical Transactions of the Royal Society of London A:
  Mathematical, Physical and Engineering Sciences}, 358(1777):3229--3246.

\bibitem[\protect\astroncite{Uzun and Hussaini}{2010}]{uzun10}
Uzun, A. and Hussaini, M.~Y. (2010).
\newblock Simulations of vortex formation around a blunt wing tip.
\newblock {\em AIAA Journal}, 48:1221--1234.

\bibitem[\protect\astroncite{Visbal}{2011}]{visbal11}
Visbal, M.~R. (2011).
\newblock Numerical investigation of deep dynamic stall of a plunging airfoil.
\newblock {\em AIAA Journal}, 49(10):2152 -- 2170.

\bibitem[\protect\astroncite{Visbal}{2014}]{visbal14}
Visbal, M.~R. (2014).
\newblock Analysis of the onset of dynamic stall using high-fidelity large-eddy
  simulations.
\newblock In {\em 52nd Aerospace Sciences Meeting, AIAA SciTech Forum}. AIAA.

\bibitem[\protect\astroncite{Visbal and Garmann}{2017}]{visbal17}
Visbal, M.~R. and Garmann, D.~J. (2017).
\newblock Numerical investigation of spanwise end effects on dynamic stall of a
  pitching naca 0012 wing.
\newblock In {\em 55th AIAA Aerospace Sciences Meeting, AIAA SciTech Forum}.
  AIAA.

\bibitem[\protect\astroncite{Yuan et~al.}{2013}]{yuan13}
Yuan, W., Poirel, D., and Wang, B. (2013).
\newblock Simulations of pitch-heave limit-cycle oscillations at a transitional
  reynolds number.
\newblock {\em AIAA Journal}, 51:1716 -- 1732.

\end{thebibliography}

\end{document}